\documentclass[fleqn,usenatbib]{mnras}

\DeclareRobustCommand{\VAN}[3]{#2}
\let\VANthebibliography\thebibliography
\def\thebibliography{\DeclareRobustCommand{\VAN}[3]{##3}\VANthebibliography}

\usepackage{graphicx}	
\usepackage{amsmath}	
\usepackage{amssymb}	
\usepackage{array}

\title[Radio observations of AT2019azh]{AT2019azh: an unusually long-lived, radio-bright thermal tidal disruption event}

 \author[Goodwin et. al.]{A. J. Goodwin$^1$, S. van Velzen$^{2,3,4}$, J. C. A. Miller-Jones$^1$, A. Mummery$^{5}$, M. F. Bietenholz$^{6,7}$,
 \newauthor{A. Wederfoort${^2}$, E. Hammerstein$^{8}$, C. Bonnerot$^{9}$, J. Hoffmann$^{1}$, and L. Yan$^{10}$}
\\
$^{1}$International Centre for Radio Astronomy Research -- Curtin University, GPO Box U1987, Perth, WA 6845, Australia \\
$^{2}$Leiden Observatory, Leiden University, PO Box 9513, 2300 RA Leiden, The Netherlands \\
$^{3}$Center for Cosmology and Particle Physics, New York University, NY 10003, USA\\
$^{4}$Department of Astronomy, University of Maryland, College Park, MD 20742, USA\\
$^{5}$ Oxford Astrophysics, Denys Wilkinson Building, Keble Road, Oxford, OX1 3RH, United Kingdom \\
$^{6}$ Department of Physics and Astronomy, York University, Toronto, M3J 1P3, Ontario, Canada\\
$^{7}$ SARAO/Hartebeesthoek Radio Observatory, PO Box 443, Krugersdorp, 1740, South Africa \\
$^{8}$ Department of Astronomy, University of Maryland, College Park, MD 20742, USA\\
$^{9}$ TAPIR, Mailcode 350-17, California Institute of Technology, Pasadena, CA 91125, USA\\
$^{10}$The Caltech Optical Observatories, California Institute of Technology, Pasadena, CA 91125, USA\\
}

\date{Accepted XXX. Received YYY; in original form ZZZ}

\pubyear{2021}

\begin{document}
\label{firstpage}
\pagerange{\pageref{firstpage}--\pageref{lastpage}}
\maketitle

\begin{abstract}
Tidal disruption events (TDEs) occur when a star is destroyed by a supermassive black hole at the center of a galaxy, temporarily increasing the accretion rate onto the black hole and producing a bright flare across the electromagnetic spectrum. 
Radio observations of TDEs trace outflows and jets that may be produced. Radio detections of the outflows from TDEs are uncommon, with only about one third of TDEs discovered to date having published radio detections. 
Here we present over two years of comprehensive, multi-radio frequency monitoring observations of the tidal disruption event AT2019azh taken with the Very Large Array (VLA) and MeerKAT radio telescopes from approximately 10 days pre-optical peak to 810 days post-optical peak.
AT2019azh shows unusual radio emission for a thermal TDE, as it brightened very slowly over two years, and showed fluctuations in the synchrotron energy index of the optically thin synchrotron emission from 450 days post-disruption. 
Based on the radio properties, we deduce that the outflow in this event is likely non-relativistic and could be explained by a spherical outflow arising from self-stream intersections, or a mildly collimated outflow from accretion onto the supermassive black hole.
This data-set provides a significant contribution to the observational database of outflows from TDEs, including the earliest radio detection of a non-relativistic TDE to date, relative to the optical discovery.  

\end{abstract}




\section{Introduction}

The central mass in a galaxy influences the dynamics and spatial distributions of stars in the inner regions. In some rare cases a supermassive black hole (SMBH) can capture and destroy a star if it passes within the radius at which the tidal shear forces on the star from the black hole exceed the star's self-gravity \citep{Rees1988}. Such tidal disruption events (TDEs) produce bright flares across the electromagnetic spectrum that are usually visible for timescales of 1--2 years, with approximately half of the debris remaining in orbits bound to the black hole, and other parts flung out on hyperbolic orbits with large velocities \citep[e.g.][]{Lacy1982,Rees1988,Evans1989,Lodato2009}. The bound stellar debris may circularise and form an accretion disk \citep[e.g.][]{Shiokawa2015,Bonnerot2020,Bonnerot2016,Liptai2019,Hayasaki2016,Mummery2020}, producing X-ray emission from the accretion stream as the material falls back towards the black hole, and optical emission possibly from the re-processing of the X-rays in the disk \citep[e.g.][]{Auchettl2017,Cannizzo1990,vanVelzen2020,Gezari2021,Strubbe2009,Metzger2016,Roth2016}. The disk circularisation time, and thus the time taken for disk formation and accretion onto the SMBH to begin, is a subject of debate \citep{Bonnerot2021}. Some models predict varying circularisation timescales depending on the physical properties of the star, SMBH, and system \citep[e.g.][]{Lu2019,Liptai2019,Hayasaki2016,Bonnerot2021}. X-ray observations of TDEs that trace accretion onto the SMBH have shown variable behaviour, with some TDEs showing bright X-ray emission early on \citep[e.g.][]{Miller2015}, and others showing delayed X-ray flares \citep[e.g.][]{Hinkle2021}.  In some cases radio emission is also observed from outflowing material.

Radio observations of TDEs trace the outflows produced by the debris from the destroyed star that may be ejected from the black hole \citep[see][for a review]{Alexander2020}, enabling detailed insight into the launching of outflows from SMBHs, the circumnuclear density, and how such outflows might provide feedback and influence the evolution of the host galaxy. 
The radio properties of the TDEs observed to date are diverse, with some exhibiting high luminosity emission ($\nu L_{\nu}>10^{40}$ erg/s)  that is well-described by a relativistic jet \citep[e.g. Swift J164449.3+573451 (Sw J1644+57),][]{Levan2011,Burrows2011,Zauderer2011,Bloom2011} and others exhibiting lower-luminosity emission ($\nu L_{\nu}<10^{40}$ erg/s) that could be described by synchrotron emission from a non-relativistic spherical or mildly collimated outflow \citep[e.g. ASASSN-14li,][]{vanVelzen2016,Alexander2016}. The time relative to the optical/X-ray flare at which the radio emission is observed could provide insight into the mechanism launching the outflow, the star that was destroyed, and the nature of its orbit around the black hole. Recently it has been suggested that delayed radio emission may be common in TDEs \citep{Horesh2021b}, based on late-time radio flares observed for the thermal TDEs ASASSN-15oi and iPTF16fnl. However, radio observations of TDEs at early times are uncommon, and the apparent lack of detected early-time radio emission could naturally result from a lack of early-time observations, as we demonstrate through early-time radio observations of the thermal TDE AT2019azh in this work.
 
Due to the diverse properties of the TDEs that have been observed in the radio to date, there is no consistent explanation for the type of outflows that may be produced in any single event. In some rare cases, as in Sw J1644+57, a relativistic jet is produced with energy $\sim10^{51}$ erg \citep{Burrows2011,Brown2017,Pasham2015,Cenko2012,Zauderer2011}. These arise in TDEs that present a non-thermal X-ray spectrum, and the radio emission can be well described by a relativistic jet model in which synchrotron emission is produced as the jet shocks the circumnuclear medium (CNM) and slows, producing emission similar to a gamma-ray burst \citep[e.g.][]{Metzger2012,Kumar2013}. In other cases, where the TDEs exhibit a thermal X-ray spectrum (as for example in ASASSN-14li and AT2019dsg), a less energetic ($E\sim10^{46}-10^{50}$\,erg), non-relativistic outflow is produced. 

There are a few possible scenarios for producing the observed properties of these non-relativistic outflows, some of which are difficult to rule out with the current set of observations. The prevalent models include a disk wind model in which the outflow is produced early on by accretion onto the SMBH, and emits synchrotron radiation as it moves through the interstellar medium around the black hole \citep[e.g.][]{Alexander2016}. Alternatively, in a similar scenario, a mildly collimated, non-relativistic jet could be produced by the accretion onto the SMBH \citep[e.g.][]{vanVelzen2016}. In this case, the jet, emitting radio emission by an internal emission mechanism, could switch on at later times and a constant injection of energy could be observed, with the energy increasing with time \citep[e.g.][]{Falcke1995}. Another possibility is a collision-induced outflow, in which the debris from the destroyed star undergoes stream-stream collisions as it circularises into an accretion disk, with a significant amount of gas becoming unbound and ejected in an approximately spherical outflow \citep{Lu2019}. Finally, the radio emission could also be produced by the unbound tidal debris stream, which would be ejected from the system with escape velocities $\sim10^4$\,km/s in a concentrated cone close to the orbital plane \citep{Krolik2016}.

Recently there has been an increase in the number of TDEs with radio detections \citep[e.g.][]{Alexander2016,vanVelzen2016,Alexander2017,Cendes2021,Horesh2021}, with the next few years expected to bring a large number of new radio observations of these unique events due to targeted radio campaigns to follow-up optical and X-ray detected events. These new observations will be crucial in characterising the mechanism behind the radio-emitting outflows that can be produced, and identifying if there is a single mechanism behind all radio outflows, or if the type of outflow differs between individual systems.

In this work we present over two years of radio monitoring observations of the thermal TDE AT2019azh. AT2019azh was first discovered on 2019 February 22 by the All-Sky Automated Survey for Supernovae (ASASSN) \citep{ASSASN_Atel} and named ASASSN-19dj. It was detected by the Zwicky Transient Facility (ZTF) on 2019 February 12 and denoted ZTFaaazdba \citep{ZTF_Atel}. The source was coincident with the nucleus of the E+A galaxy KUG 0810+227, with a redshift of $z=0.022$ (luminosity distance of 96\,Mpc). Spectra obtained by the Nordic Optical Telescope Unbiased Transient Survey (NUTS) on 2019 February 22 \citep{NUTS_Atel}, ePESSTO on 2019 February 25 \citep{ePESSTO_Atel}, and the Spectral Energy Distribution Machine mounted on the Palomar 60-in telescope on 2019 February 24 and March 10 \citep{ZTF_Atel} all revealed a blue, featureless spectrum with narrow emission and Balmer absorption features associated with the host galaxy. The event was also detected in X-ray and UV with the Neils Gehrels Swift Observatory and was found to have a high X-ray blackbody temperature of $kT=0.06$\,eV assuming a thermal spectrum \citep{ZTF_Atel}. The combination of optical spectral properties, high blackbody temperature, location at center of host galaxy, and lack of spectroscopic AGN or supernova features led \citet{ZTF_Atel} to classify the source as a tidal disruption event. The first reported radio detection of the event was by \citet{PerezcollaboratorsATel} with the electronic Multi-Element Remotely Linked Interferometer Network (e-MERLIN) at 5\,GHz on 2019 May 21 and 2019 June 11. In this work we present an earlier radio detection, on 2019 March 9.

\citet{Hinkle2021} analysed the optical, UV, and X-ray observations of AT2019azh from 30\,d before to $\sim$300\,d after the optical peak. During the first 200\,d there was very low-level X-ray emission with a harder spectral index, and strong optical/UV emission that evolved as expected for a thermal TDE. The X-rays brightened by a factor of 30--100 approximately 250\,d after discovery, and the X-ray spectrum became softer. The optical/UV flare was observed to begin on MJD 58528 (2019 Feb 14) and peaked on approximately MJD 58560 \citep[2019 Mar 18,][]{Hinkle2021}. From a power-law fit of the optical rise observed by ASASSN, \citet{Hinkle2021} inferred that the time of first light for the event was MJD 58522 (2019 Feb 8). 

\citet{Liu2019} found X-ray flaring episodes during the early times, which were temporally uncorrelated with the optical/UV emission. They deduced that the optical and X-ray data could be explained by a two-process scenario in which the early emission in UV/optical is explained by emission from debris stream-stream collisions as the bound debris is becoming circularised, and the low-level early X-ray emission is due to a low-mass accretion disk forming during this time. \citet{Liu2019} explain the late X-ray brightening as due to the major body of the disk forming after circularisation of the bound stellar debris. 
However, \citet{Hinkle2021} concluded that the late-time X-ray brightening was a consequence of an increase in the area of the X-ray emitting region via the blackbody radius, while the short term X-ray variability was due to changes in the X-ray temperature. In this scenario, the late-time X-ray brightening is not due to delayed accretion disk formation, but rather an expansion of the X-ray emitting region. However, \citet{Mummery2021b} recently showed that the X-ray blackbody radius is not a good measure of length scales in a TDE system, implying that the change in blackbody radius may not be caused by a change in the disk radius. The nature of AT2019azh is thus a subject of debate, with evidence both for and against a delayed accretion scenario to explain the multiwavelength emission from the event.

In this work we present 13 epochs of radio observations of AT2019azh taken with the Very Large Array (VLA) and MeerKAT beginning 2019 March 9 (before the optical peak) and spanning until 2021 June 5. These radio observations enable further insight into the nature of the TDE, including the disk formation and launching of the outflow.
The paper is outlined as follows: in Section \ref{sec:Observations} we describe the radio observations and data processing. In Section \ref{sec:Results} we present the radio spectral observations and synchrotron emission fits. In Section \ref{sec:Modeling} we describe the modelling of the radio emission to predict physical properties of the outflow. In Section \ref{sec:multiwavelength_modelling} we present a more detailed accretion-disk model of the multiwavelength observations. In Section \ref{sec:Discussion} we discuss the implications of these results, the possible nature of the outflow in AT2019azh, and relate the outflow properties to those of other TDEs. Finally in Section \ref{sec:Conclusion} we provide a summary of our results and concluding remarks. 

\section{Observations}\label{sec:Observations}

\subsection{VLA observations}
We obtained radio observations of AT2019azh with the NRAO's Karl G. Janksy Very Large Array (VLA) spanning from 2019 March 9 to 2021 February 26 across 300\,MHz--24\,GHz (P- to K-band; program IDs 19A-395 and 20A-392). In our first observation, on 2019 March 9, we observed the optical position of the source (RA, Dec) 08:13:16.945, +22:38:54.03 and detected faint radio emission at 10\,GHz with a flux density of $150\pm12$\,$\mathrm{\mu Jy}$. The position of this radio emission was (RA, Dec) 8:13:16.95, +22.38.54.02 with a positional accuracy of 1\,arcsecond, coincident with the optical position. We subsequently triggered follow-up observations over a broader frequency range, and continued to monitor the source evolution over the following two and a half years, taking 13 epochs of observations in total. The observations are summarised in Table \ref{tab:radio_obs} (flux densities and frequencies are available in the online machine-readable format of the table).

All data were reduced in the Common Astronomy Software Application package \citep[\texttt{CASA 5.6,}][]{McMullin2007} using standard procedures. Where possible, we calibrated the data using the VLA calibration pipeline available in \texttt{CASA}. In all observations 3C 147 was used as the flux density calibrator. For phase calibration we used ICRF J082324.7+222303 for 12--26\,GHz (K- and Ku-band), ICRF J083216.0+183212 for 4--12\,GHz (X- and C-band); ICRF J084205.0+183540 for 1--4\,GHz (S- and L-band); and PKS J0801+1414 for 0.23--0.47\,GHz (P-band). The P-band data were reduced manually using standard procedures in \texttt{CASA}, including phase and amplitude self-calibration. Images of the target field of view were created using the \texttt{CASA} tasks \texttt{CLEAN} or \texttt{TCLEAN} (for epochs post April 2019) for all bands except P-band, where we used the \texttt{WSCLEAN} 
($w$-stacking CLEAN) imager \citep{offringa-wsclean-2014, OffringaS2017}. The source flux density was measured in the image plane, by fitting an elliptical Gaussian fixed to the size of the synthesized beam using the \texttt{CASA} task \texttt{IMFIT}. The errors associated with the measured flux densities include a statistical uncertainty and a systematic one due to the uncertainty in the flux-density bootstrapping, estimated at 5\%. Where enough bandwidth was available, we split the L-, C-, and S-band data into four sub-bands when imaging, and the X-band data into two sub-bands. The source was detected at a 4-$\sigma$ confidence level in the P-band observations, so we did not split these data into sub-bands. 

\subsection{MeerKAT observations}
We also observed AT2019azh with MeerKAT on four occasions between 2019
November 29 and 2020 November 14.  We used the 4K (4096-channel)
wideband continuum mode and observed with bandwidth of 856~MHz around
a central frequency of 1.28~GHz.  Each observation was about 2~h long
in total, except for that on 2020 Nov.\ 14, which was only 1~h long.

The data were reduced using the OxKAT scripts \citep{Heywood2020}.  We
used PKS J0408$-$6544 (QSO B0408$-$65) to set the flux density scale
and calibrate the bandpass, and ICRF J084205.0+183540
as a secondary calibrator.  The final images were made using the \texttt{WSCLEAN} imager \citep{offringa-wsclean-2014, OffringaS2017}, and resolved
into 8 layers in frequency.  \texttt{WSCLEAN} deconvolves the 8 frequency layers
together by fitting a polynomial in frequency to the brightness in the
8 frequency-layers.  Our flux densities include both the
statistical uncertainty and a systematic one due to the uncertainty
in the flux-density bootstrapping, estimated at 5\%.
\\

To ensure no systematic offset between epochs and instruments, in Appendix \ref{sec:appendix1} we present an analysis of flux density measurements of three background sources for 9 epochs of VLA data and 4 epochs of MeerKAT data. We found no significant systematic offset between the two instruments, and found flux densities between VLA epochs were consistent to within $\sim10\%$. The flux scale obtained through calibration of the VLA data is consistent across epochs to within a few percent, indicating that the flux density fluctuations we infer between epochs are larger than that expected through calibration differences alone. However, there is no systematic frequency dependence for these inter-epoch flux density variations, and these differences between epochs could be due to intrinsic variability of the background sources, which are expected to be variable at some level.

 \begin{table}
 \label{tab:radio_obs}
 \caption{Radio observations of AT2019azh taken with the VLA and MeerKAT.}
 \newcolumntype{~}{!{\vrule width 0.5pt}}
 \begin{tabular}{p{0.4cm}p{1.5cm}p{0.4cm}p{0.5cm}p{1cm}p{1.5cm}}
        \hline
        Epoch & Date (UTC) & $\delta$t (d) &  Array& Config. & Frequency Bands \\
        \hline
        \hline
        1 & 2019-03-09 1:10 & 29 & VLA & B & X \\
 
        2 & 2019-04-14 2:34 & 65 & VLA & B & K, X, C  \\

        3 & 2019-05-12 20:39 & 94 & VLA & B & Ku, X, C \\

        4 & 2019-05-20 23:51 & 102 & VLA & B & C, S, L \\
 
        5 & 2019-06-19 23:40 & 132 & VLA & B & X, C, S, L \\

        6 & 2019-08-09 19:06 & 183& VLA & A & X, C, S, L \\

        7 & 2019-10-19 15:04 & 254& VLA & A & X, S, L \\

        7 & 2019-11-25 1:02 &  & MeerKAT &  & L  \\

        8 & 2019-11-30 12:32 & 296 & VLA & D & X, C, S  \\
        8 & 2020-01-29 21:50 &  & MeerKAT &  & L \\

        9 & 2020-01-24 9:27 & 350 & VLA & D & X, S\\
        9 & 2020-05-05 15:08 &  & MeerKAT &  & L \\

        10 & 2020-05-11 22:19 & 459 & VLA & C & X, C, S \\

       10 & 2020-11-14 4:40 &  & MeerKAT &  & L \\

        11 & 2020-12-05 08:02 & 666 & VLA & bnA->A & X, C, S, L, P\\

        12 & 2021-02-26 07:03 & 749 & VLA & A & X, C, S, L, P \\

        13 & 2021-05-06 20:02 & 849 & VLA & D->C & X, C, S, L\\

        \hline
    \end{tabular}
    \\
    \textbf{Notes:} $\delta t$ is measured with reference to the estimated outflow launch date, $t_0=$ MJD\,58522. For the frequency bands: P=0.23-0.47\,GHz, L=1--2\,GHz, S=2--4\,GHz, C=4--8\,GHz, X=8--12\,GHz, Ku=12--18\,GHz, and K=18--26.5\,GHz. A complete version of this table including flux density measurements is available in machine-readable format from \textcolor{red}{url to be inserted on publication}. 
\end{table}

\subsection{Multiwavelength observations}

We obtained forced point-spread function fitting (PSF) photometry of AT2019azh from the public ZTF MSIP data through the ZTF forced-photometry service \citep{Masci19}. We filtered the resulting optical light curves for observations impacted by bad pixels, and required thresholds for the signal-to-noise of the observations, seeing, the sigma-per-pixel in the input science image, and several parameters relating to the photometric and astrometric calibrators.

The majority of the \textit{Swift} UVOT observations were published in \citet{vanVelzen21}. Here, we include new observations taken after the publication of that work. We used the \texttt{uvotsource} package to analyze the \textit{Swift} UVOT photometry and the resulting UV data have been host galaxy subtracted. We also include NICER and XMM-Newton observations reported in \citet{Hinkle2021}.

\section{Radio lightcurve and spectra}\label{sec:Results}

\begin{figure*}
    \centering
    \includegraphics[width=\columnwidth]{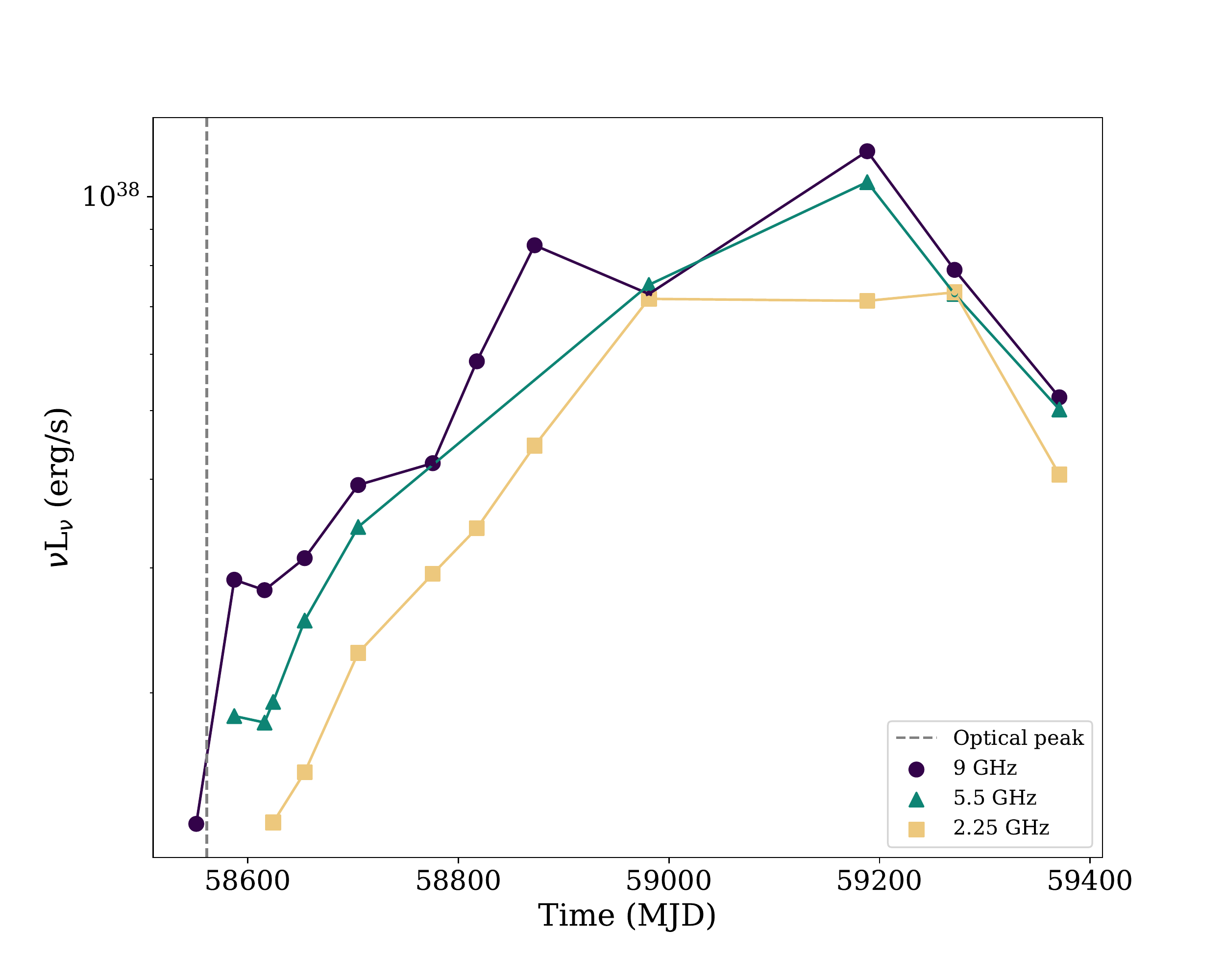}
    \includegraphics[width=\columnwidth]{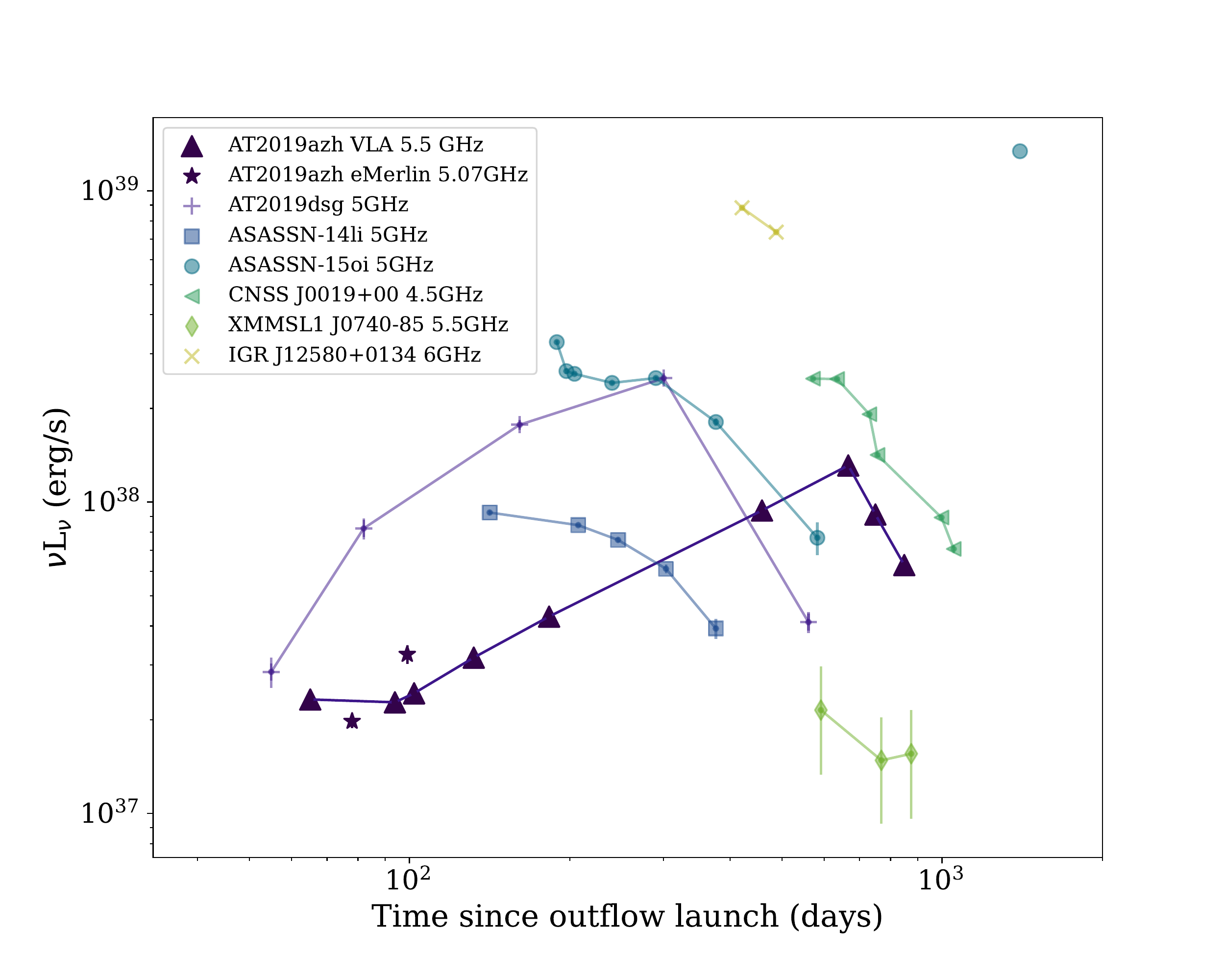}
    \caption{\textit{Left}: The luminosity of AT2019azh inferred from VLA observations at 9 (purple circles), 5.5 (green triangles) and 2.25 (yellow squares) GHz. \textit{Right:} The luminosity of AT2019azh at 5.5\,GHz inferred from VLA monitoring (purple triangles) and e-MERLIN observations reported in \citet{PerezcollaboratorsATel} (purple stars). For comparison, the $\approx5$\,GHz luminosity of 6 other radio detected thermal TDEs are shown. TDE data are from \citet{Alexander2016,vanVelzen2016} (ASSASN-14li), \citet{Cendes2021,Stein2021} (AT2019dsg), \citet{Horesh2021} (ASASSN-15oi), \citet{Anderson2020} (CNSS J0019+00), \citet{Alexander2017} (XMSSL J0740-85), and \citet{Irwin2015} (IGR J12580+0134). All luminosities are plotted with reference to the approximate inferred outflow launch date, or the inferred optical first-light if no estimate of the launch date is available.}
    \label{fig:lightcurve}
\end{figure*}

The 2.25, 5, and 9 GHz VLA lightcurves for AT2019azh are plotted in Figure \ref{fig:lightcurve}, as well as a comparison of the 5\,GHz lightcurve with other thermal TDE lightcurves. The radio emission from AT2019azh rose relatively slowly at all radio wavelengths until approximately 625\,d post optical discovery, at which time the higher frequency ($>4$\,GHz) emission started to decrease while the 2\,GHz emission remained relatively constant. Such a slow rise in the radio relative to the optical peak, which occurred around the time of our first radio detection, places AT2019azh in the slow-rising thermal TDE population (Figure \ref{fig:lightcurve}). In contrast, some thermal TDEs have been observed to begin fading in the radio soon after the optical peak \citep[e.g.][]{Alexander2016,Horesh2021}. 

The 5\,GHz luminosity of AT2019azh increases approximately linearly with time, similar to that of the relativistic event Sw J1644+57. However, Sw J1644+57 rose to a peak within $\sim$100\,d \citep{Eftekhari2018}, whereas AT2019azh took over $\sim$600\,d. We note that AT2019azh was detected in the radio significantly earlier relative to the optical peak than the other thermal TDEs, and a similar slow rise cannot be ruled out for ASSASN-14li, CNSS J0019+00, or XMMSL1 J0740-85. The rise observed for AT2019azh is significantly different than those of ASASSN-15oi, which had early radio non-detections \citep{Horesh2021}), and AT2019dsg, which rose to a peak over $<350$\,d with $L\propto t^{2.5}$ \citep{Stein2021}. 

The luminosity of AT2019azh is now sharply decreasing, and similar to the fading rates of AT2019dsg, ASASSN-14li, CNSS J0019+00, and ASASSN-15oi (Figure \ref{fig:lightcurve}).

\subsection{Spectral fitting}

We fit the observed radio spectrum at each epoch using the synchrotron spectrum model described in \citet{Granot2002}. The flux density of the synchrotron emission spectrum, assuming $\nu_{\rm m} < \nu_{\rm a} < \nu_{\rm c}$  (where $\nu_{\rm m}$ is the synchrotron minimum frequency, $\nu_{\rm a}$ is the synchrotron self-absorption frequency, and $\nu_{\rm c}$ is the synchrotron cooling frequency), is described by 
\begin{equation}
\begin{aligned}
    F_{\nu, \mathrm{synch}} = F_{\nu,\mathrm{ext}} \left[\left(\frac{\nu}{\nu_{\rm m}}\right)^2 \exp(-s_1\left(\frac{\nu}{\nu_{\rm m}}\right)^{2/3}) + \left(\frac{\nu}{\nu_{\rm m}}\right)^{5/2}\right] \times \\
    \left[1 + \left(\frac{\nu}{\nu_{\rm a}}\right)^{s_2(\beta_1 - \beta_2)}\right]^{-1/s_2}
    \end{aligned}
\end{equation}
where $\nu$ is the frequency, $F_{\nu,\mathrm{ext}}$ is the normalisation, $s_1 = 3.63p-1.60$, $s_2 = 1.25-0.18p$, $\beta_1 = \frac{5}{2}$, $\beta_2 = \frac{1-p}{2}$, and $p$ is the synchrotron energy index.

Archival Faint Images of the Radio Sky at Twenty-cm (FIRST) survey observations from January 1996 of the host galaxy of AT2019azh show no detection at 1.4\,GHz, with a 3\,$\sigma$ upper limit of 0.41\,mJy \citep{Becker1995}. These observations place a strong upper bound on the host galaxy contribution to the radio emission and make recent AGN activity in the galaxy unlikely. Whilst \citet{Hinkle2021} were also able to rule out strong AGN activity based on optical properties of the host galaxy, they found that optical and X-ray observations cannot rule out the presence of a low-luminosity AGN in the host galaxy, KUG 0810+227. In order to account for the possibility of some low-level contribution from the host galaxy to the observed radio flux densities of the outflow, we add a host component to the spectral fitting that is described by
\begin{equation}\label{eq:hostflux}
    F_{\nu, \mathrm{host}} = F_{0} \left( \frac{\nu}{1.4\,\mathrm{GHz}} \right)^{\alpha_0},
\end{equation}
where $F_0$ is the flux density measured at 1.4\,GHz ($F_0<0.41$\,mJy) and $\alpha_0$ is the spectral index of the host galaxy. The total observed flux is then given by
\begin{equation}\label{eq:spectralfit}
    F_{\nu,\mathrm{total}} = F_{\nu, \mathrm{host}} + F_{\nu, \mathrm{synch}}.
\end{equation}

\begin{figure*}
    \centering
    \includegraphics[width=\textwidth]{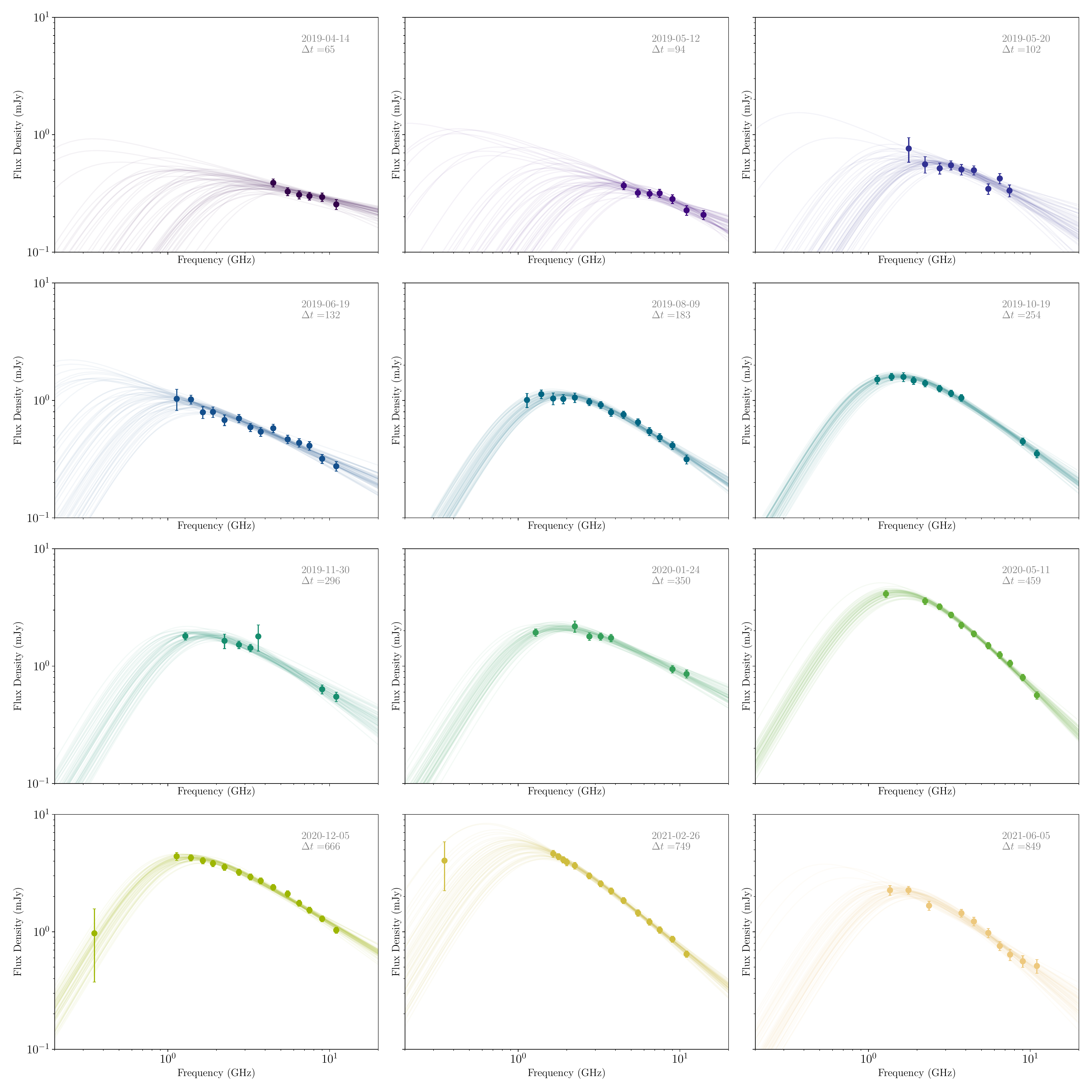}
    \caption{MCMC spectral fits (solid lines) of 12 epochs of radio observations (scatter points) of AT2019azh using the combined VLA and MeerKAT data. 50 random samples from the MCMC chains are plotted after discarding the first 1000 steps for burn-in. Note that the peak flux density and frequency of the first four epochs are not well-constrained due to the lack of low-frequency coverage.}
    \label{fig:spectral_fits}
\end{figure*}

We fit the spectra for all epochs using a Python implementation of Markov Chain Monte Carlo (MCMC), \texttt{emcee} \citep{emcee}.  We use a Gaussian likelihood function where the variance is underestimated by some fractional amount $f$. We assume flat prior distributions for all parameters, and allow $p$ to fall in the range $2.5-4.0$. Whilst it is common practice to fix $p$ between epochs \citep[e.g.][]{Alexander2016,Cendes2021}, there is evidence in the observations that the spectral slope is not constant for this event (see Figure \ref{fig:spectral_fits}), so we do not fix $p$ in our modelling. To constrain the host flux density and spectral index (Equation \ref{eq:hostflux}), we first ran a MCMC fit for epoch 11 (2020 Dec 12) only, where the synchrotron spectrum is very well constrained by the observations, ensuring that $F_{0}<0.41$\,mJy and $-2<\alpha_{0}<2$. We found the best solution for $F_{0}=0.175$\,mJy and $\alpha_0=-0.84$. 
Next we fit the total flux density using the determined host contribution as a function of frequency for the energy index, $p$, the flux normalisation, $F_{\nu,\mathrm{ext}}$, the minimum frequency, $\nu_{\rm m}$, and the self absorption frequency, $\nu_{\rm a}$ for all epochs using Equation \ref{eq:spectralfit}. For the firs three epochs the peak frequencies and flux densities are not well-constrained due to the lack of low-frequency radio coverage. The MCMC spectral fitting results for the synchrotron self-absorption break and peak flux density for these epochs are dependent on the choice of prior. Thus, for these epochs we provide upper and lower limits respectively for $\nu_{\mathrm{a}}$ and $F_{\mathrm{peak}}$. 

\begin{figure*}
    \centering
    \includegraphics[width=\textwidth]{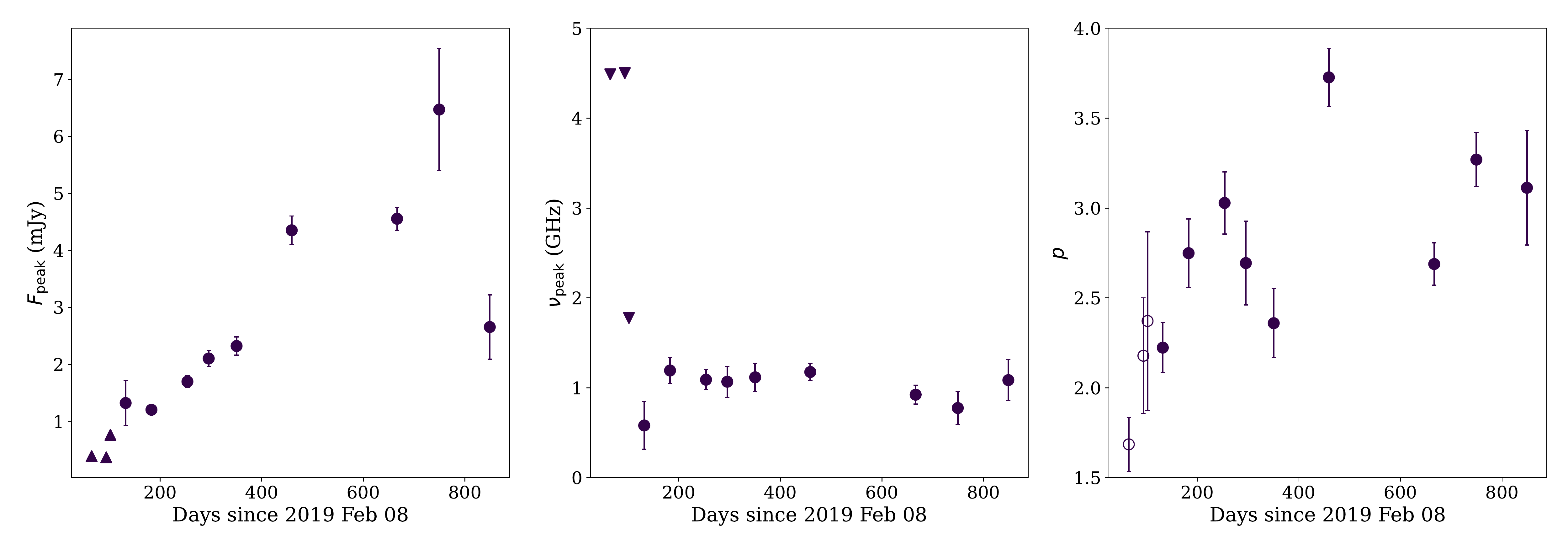}
    \caption{Spectral fit properties of the radio emission inferred using Equation \ref{eq:spectralfit} and shown in Figure \ref{fig:spectral_fits}. Upper and lower limits (triangles) are given for the epochs where the peak flux and frequency are not well-constrained by the radio observations and all error-bars represent the $1\sigma$ confidence intervals from the MCMC fitting.  The peak flux density increases approximately linearly with time, excepting the final epoch, which showed a significant drop in peak flux density. The peak frequency is approximately constant for all well-constrained epochs, with a slight downwards trend after the radio lightcurve peak ($t>600$\,d). The energy index, $p$, shows significant fluctuations after 400\,d.}
    \label{fig:spectralfitparams}
\end{figure*}

The spectral fits for each epoch are plotted in Figure \ref{fig:spectral_fits}, and the best fit peak flux densities and frequencies from the spectral fits are reported in Table \ref{tab:outflowproperties} and plotted in Figure \ref{fig:spectralfitparams}. For the epochs where the peak of the synchrotron spectrum is well constrained, we find the peak frequency, $\nu_{\mathrm{peak}}$, remained relatively constant at $\nu_{\mathrm{peak}}=1.1\pm0.3$\,GHz, with a slight downwards trend over the 800\,d spanned by our observations, whilst the peak flux density, $F_{\mathrm{peak}}$, increased approximately linearly with time and only showed signs of decreasing in the final epoch, 820\,d post-disruption. The index of the electron energy distribution, $p$, remains roughly constant at $p\approx2.7\pm0.2$, similar to that of other thermal events \citep[e.g.][]{Alexander2016,Cendes2021,Stein2021,Horesh2021} \textit{excepting} three epochs at 254, 459, and 749\,d post disruption at which the energy index shows a significant steepening to $p\approx$3, 3.7, and 3.3 respectively. We note that the slight spectral steepening at 255\,d is likely not real based on an analysis of the flux density of background sources in the field for the different epochs presented in Appendix \ref{sec:appendix1} and that the significance of the change in $p$ is 3-$\sigma$ and 2-$\sigma$ respectively for the other two epochs. We found no evidence in the background source measurements for inconsistent calibration with frequency for the two other epochs where we observed a steepening, although there are small systematic flux density offsets between most epochs (see Appendix \ref{sec:appendix1}), which we account for with the added systematic uncertainty of $5\%$ to the flux densities. We thus conclude that the steepenings we observed are real for May 2020 (epoch 10) and February 2021 (epoch 12), and are not artefacts of inconsistent calibration over the frequency ranges for those epochs. We note that the 10\% flux density offsets between epochs on the background sources suggests a possible systematic uncertainty of 10\% in the flux density calibration.  Such a systematic uncertainty would affect our peak flux density values, however, the uncertainty of the peak flux density is dominated by the peak not being well constrained in many of the epochs due to the paucity of the data at low frequencies. 

\section{Modelling of the radio emission}\label{sec:Modeling}

We model the radio emission from the outflow using the standard synchrotron emission model outlined in \citet{BarniolDuran2013}, in which the ambient electrons are accelerated into a power-law distribution by the blastwave from the outflow, $N(\gamma)\propto \gamma^{-p}$, where $\gamma$ is the electron Lorentz factor ($\gamma\geq \gamma_{\rm m}$, where $\gamma_{\rm m}$ is the minimum Lorentz factor) and $p$ is the synchrotron energy index. We assume equipartition between the electron and magnetic field energy densities in order to derive the equipartition energy and radius. Although the emitting region may not be in equipartition, we can derive estimates of the physical system parameters by parameterising the deviation from equipartition and accounting for its effect. This approach allows us to estimate key physical quantities such as the ambient electron density, magnetic field strength, mass of the emitting region, and velocity of the ejecta. We assume the fraction of the total energy in the magnetic field is 0.1$\%$, $\epsilon_B=10^{-3}$, based on observations of other TDEs and supernovae \citep[e.g.][]{Eftekhari2018,Horesh2013}. We assume that 10$\%$ of the total energy is carried by the electrons, $\epsilon_e=0.1$, given that electrons are typically accelerated much less efficiently than protons in astrophysical accelerators \citep[e.g.][]{Morlino2012}. We find no evidence for a relativistic outflow (see Section \ref{sec:outflowmodels}), and assume the outflow is non-relativistic (i.e. the bulk Lorentz factor, $\Gamma=1$), and that the peak of the radio spectrum is associated with the synchrotron self-absorption frequency (i.e. $\nu_{\rm a}$ = $\nu_{\rm p}$). We model two different geometries, one where the emitting region is approximately spherical (with geometric factors\footnote{The geometric factors, defined in \citet{BarniolDuran2013}, are given by $f_{\mathrm{A}}=A/(\pi R^2/\Gamma^2)$ and $f_{\mathrm{v}} = V/(\pi R^3/\Gamma^4)$, for an outflow with area, $A$, volume, $V$, and distance from the origin of the outflow, $R$.} $f_{\mathrm{A}}=1$ and $f_{\mathrm{V}}=4/3$), and one where the emitting region is conical (with geometric factors $f_{\mathrm{A}}=0.13$ and $f_{\mathrm{V}}=1.15$), corresponding to a mildly collimated outflow with a half-opening angle of 30\,degrees. 

In the Newtonian regime, the equipartition energy, corresponding to the minimum total energy in the observed region, assuming $\nu_{\rm a} > \nu_{\rm m}$, is given by \citep{BarniolDuran2013}
\begin{equation}\label{eq:E_eq}
\begin{aligned}
\begin{split}
    E_{\mathrm{eq}} = 1.3\times10^{48}\, 21.8^{-\frac{2(p+1)}{13+2p}}
    (525^{(p-1)} \chi_{\rm e}^{(2-p)})^{\frac{11}{13+2p}}\\
    F_{\mathrm{peak,mJy}}^{\frac{14+3p}{13+2p}} \left(\frac{d}{10^{28}\,\rm{cm}}\right)^{\frac{2(3p+14)}{13+2p}} \left(\frac{\nu_{\mathrm{peak}}}{10\,\rm{GHz}}\right)^{-1} (1+z)^{\frac{-27+5p}{13+2p}}\\
    f_{\mathrm{A}}^{-\frac{3(p+1)}{13+2p}}
    f_{\mathrm{V}}^{\frac{2(p+1)}{13+2p}} 4^{\frac{11}{13+2p}} \xi^{\frac{11}{13+2p}}\quad\rm{erg}, 
    \end{split}
\end{aligned}
\end{equation}
where $d$ is the distance from the observer, $z$ is the redshift, $\chi_{\rm e} = \left(\frac{p-2}{p-1}\right) \epsilon_{\rm e} \frac{m_{\rm p}}{m_{\rm e}}$ ($m_{\rm e}$ is the electron mass and $m_{\rm p}$ is the proton mass), or $\chi_{\rm e}=2$ if $\Gamma=1$ (Newtonian case), and $\xi = 1 + \frac{1}{\epsilon_{\rm e}}$.

The equipartition radius is given by 
\begin{equation}\label{eq:R_eq}
\begin{aligned}
\begin{split}
    R_{\mathrm{eq}} = 1\times10^{17} (21.8 (525^{(p-1)})^{\frac{1}{13+2p}}
    \chi_{\rm e}^{\frac{2-p}{13+2p}}
    F_{\mathrm{peak}}^{\frac{6+p}{13+2p}} \left(\frac{d}{10^{28}\,\rm{cm}}\right)^{\frac{2(p+6)}{13+2p}}\\ \left(\frac{\nu_{\mathrm{peak}}}{10\,\rm{GHz}}\right)^{-1}
    (1+z)^{-\frac{19+3p}{13+2p}}
    f_{\mathrm{A}}^{-\frac{5+p}{13+2p}} f_{\mathrm{V}}^{-\frac{1}{13+2p}} 4^{\frac{1}{13+2p}} \xi^{\frac{1}{13+2p}}\quad\rm{cm}. 
\end{split}
\end{aligned}
\end{equation}

To infer the physical system parameters we first correct the energy and radius for the system being out of equipartition using the following assumptions
\begin{equation}\label{eq:R}
\begin{aligned}
R = R_{\mathrm{eq}} \epsilon^{(1/17)}
\end{aligned}
\end{equation}
\begin{equation}\label{eq:E}
\begin{aligned}
E = E_{\mathrm{eq}}\left((11/17)\epsilon^{(-6/17)} + (6/17)\epsilon^{(11/17)}\right),
\end{aligned}
\end{equation}
where $\epsilon=\frac{\epsilon_B}{\epsilon_e}\frac{11}{6}$, i.e. the total energy is minimised with respect to $R$ at $R_{\mathrm{eq}}$ when the energy in the magnetic field is $6/11$ the energy in the electrons, so the deviation from equipartition is parameterised by $\epsilon$ \citep{BarniolDuran2013}.

Using the corrected radius, $R$ and corrected energy, $E$, we then calculate the total number of electrons in the observed region ($N_{\mathrm{e}}$), ambient electron density ($n_e$), magnetic field ($B$), mass of the emitting region ($M_{\mathrm{ej}}$), and outflow velocity ($\beta=v/c$), using Equations \ref{eq:Ne}--\ref{eq:Mej} \citep{BarniolDuran2013}, as
\begin{equation}\label{eq:Ne}
\begin{aligned}
\begin{split}
    N_{\mathrm{e}} = 4\times10^{54}
    F_{\mathrm{peak,mJy}}^3 
    \left(\frac{d}{10^{28}\,\rm{cm}}\right)^{6} \left(\frac{\nu_{\mathrm{peak}}}{10\,\rm{GHz}}\right)^{-5} (1+z)^{-8}\\
    f_{\mathrm{A}}^{-2} 
    \left(\frac{R}{10^{17}\,\rm{cm}}\right)^{-4} \left(\frac{\gamma_{\rm m}}{\gamma_{\rm a}}\right)^{1-p}\quad\rm{electrons},
\end{split}
\end{aligned}
\end{equation}
where $\left(\frac{\gamma_{\rm m}}{\gamma_{\rm a}}\right)^{1-p}$ is a correction factor to account for the regime where $\nu_{\rm m} < \nu_{\rm a}$ and the extra factor of 4 arises due to the Newtonian correction; $\gamma_{\rm m}=2$, and $\gamma_{\rm a}$ is given by
\begin{equation}
    \gamma_{\rm a} = 525\,  F_{\mathrm{peak}} 
    \left(\frac{d}{10^{28}\,\rm{cm}}\right)^2 \left(\frac{\nu_{\mathrm{peak}}}{10\,\rm{GHz}}\right)^{-2} (1+z)^{-3} \frac{1}{{f_{\mathrm{A}}} \left(\frac{R}{10^{17}\,\rm{cm}}\right)^{2}}.
\end{equation}

The ambient electron density is then inferred via 
\begin{equation}\label{eq:n}
n = N_{\mathrm{e}}/V
\end{equation}

We determine the velocity of the outflow, $\beta$, by rearranging Equation 22 from \citet{BarniolDuran2013}, where the observer time, $t$ is given by

\begin{equation}\label{eq:vej}
    t = \frac{R (1-\beta)(1+z)}{\beta c},
\end{equation}
where we set the time $t$ relative to the approximate launch date of the outflow, MJD 58522; inferred based on a linear fit to the predicted radius and estimated optical time of first light (see below). The magnetic field is given by

\begin{equation}\label{eq:B}
\begin{aligned}
    B = 1.3\times10^{-2}
    F_{\mathrm{peak,mJy}}^{-2} 
    \left(\frac{d}{10^{28}\,\rm{cm}}\right)^{-4} \left(\frac{\nu_{\mathrm{peak}}}{10\,\rm{GHz}}\right)^{5}
    (1+z)^{7}\\
    f_{\mathrm{A}}^{2} \left(\frac{R}{10^{17}\,\rm{cm}}\right)^{4}\quad\rm{G},
\end{aligned}
\end{equation}
and the approximate mass in the emitting region of the ejecta is given by
\begin{equation}\label{eq:Mej}
\begin{aligned}
    M_{\mathrm{ej}} = \frac{2E}{\beta^2},
\end{aligned}
\end{equation}
noting that this is a lower limit on the mass in the emitting region of the outflow due to the energy estimate from equipartition also being a lower limit on the energy. 

\begin{figure*}
    \centering
    \includegraphics[width=\textwidth]{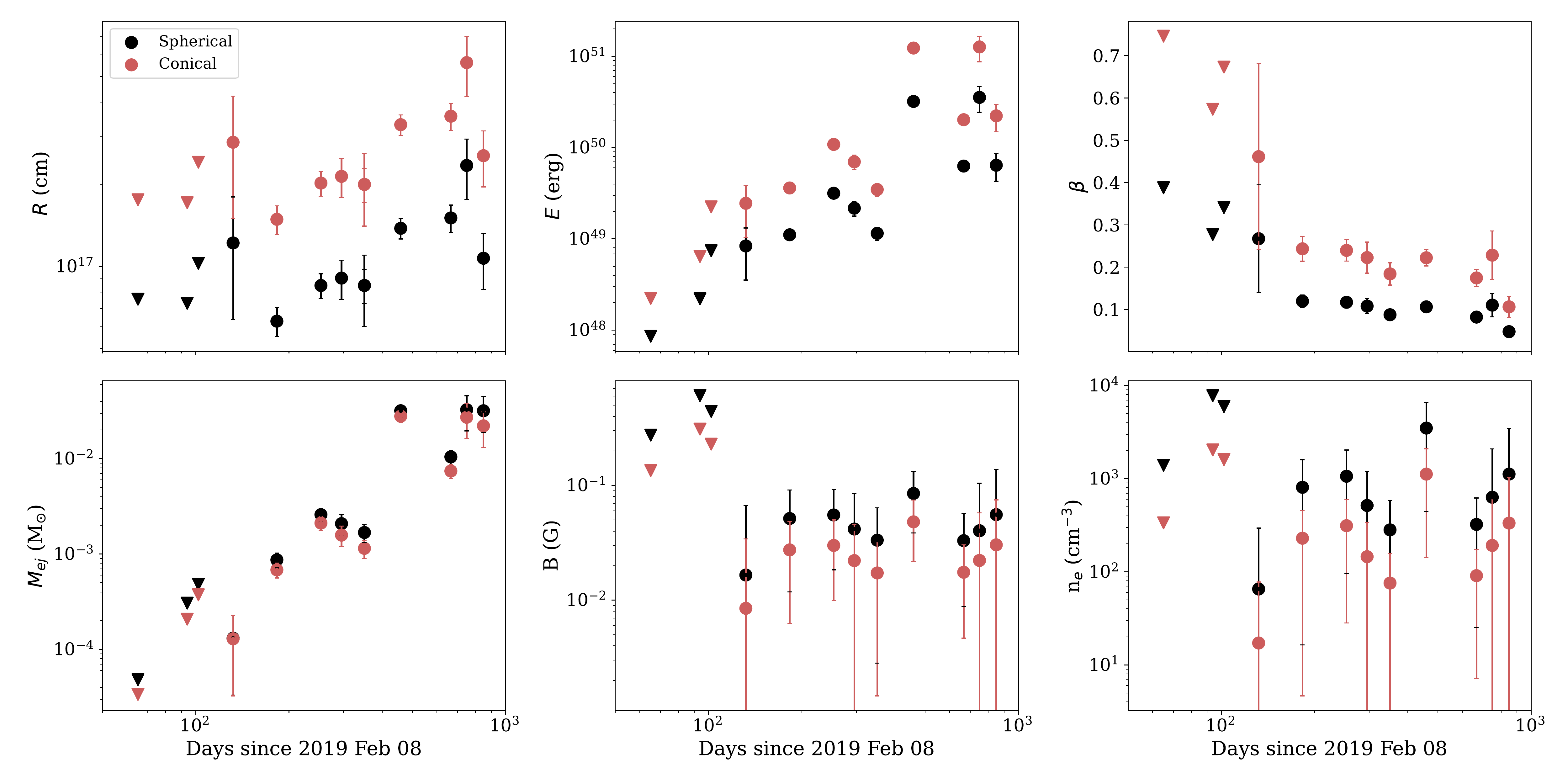}
    \caption{Physical properties of the outflow produced in the TDE AT2019azh inferred from an equipartition analysis of the peak radio flux and frequency assuming a spherical, non-relativistic outflow (black) and a conical, non-relativistic outflow (red). Upper limits (triangles) are given for the epochs where the peak flux and frequency are not well-constrained by the radio observations. The energy and radius increase approximately linearly with time until the final epoch. The velocity and magnetic field remain approximately constant over the $\sim800$\,d spanned by our observations.}
    \label{fig:blastwavemodel}
\end{figure*}

The physical outflow properties as predicted by these equations are listed in Table \ref{tab:outflowproperties} and plotted in Figure \ref{fig:blastwavemodel}. All uncertainties reported correspond to the 1$\sigma$ uncertainty obtained through propagating the uncertainty of $\nu_{\mathrm{peak}}$, $F_{\mathrm{peak}}$ and $p$ obtained through the MCMC modelling of the observed spectra.

The derived radius of the outflow increases with time, following the relation $R\propto t^{0.65}$ (reduced $\chi^2$=1.79), or $R\propto t$ (reduced $\chi^2$=2.67). We thus deduce that the outflow is roughly undergoing free expansion, with the velocity remaining approximately constant at $\beta\approx0.2\pm0.1$ (conical) or $\beta\approx0.1\pm0.06$ (spherical) until the final epoch in which the velocity shows a slight decrease. We note that the powerlaw fit to $R$, indicating a decelerating outflow, is statistically preferred to the constant velocity case. However, given the underlying assumptions of the radius calculation, that the synchrotron peak flux density and frequency were not resolved by the observations for the first few epochs, and that outflows from other thermal TDEs have all been observed to undergo approximately free-expansion at early times \citep[e.g.][]{Alexander2016,Cendes2021}, in the sections that follow we operate under the assumption that the outflow was freely expanding with $\sim$constant velocity until at least the radio lightcurve peak ($t\approx650$\,d). 

The derived energy of the outflow increases approximately linearly with time for all observations, but also seems to show statistically significant non-uniform fluctuations with time. The magnetic field shows a slight decrease with time and the inferred mass in the emitting region of the outflow also increases with time (based on the energy prediction).

\begin{table*} 
	\centering
	\caption{Predicted outflow properties of AT2019azh based on MCMC fitting of the observed radio spectrum and a synchrotron equipartition analysis.}
	\label{tab:outflowproperties}
	\begin{tabular}{lcccccccccc} 
		\hline
		 & Epoch & $\delta t$ (d) & $\nu_{\mathrm{peak}}$ (GHz) & $F_{\rm peak}$ (mJy) & $p$ & log$_{10}$\,$R$ (cm) & log$_{10}$\,$E$ (erg) & $\beta$ & log$_{10}$\,$B$ (G) & log$_{10}$\,$n_e$ \\
		 &&&&&&&&&&(cm$^{-3}$)\\
		\hline
		\hline

& 2$^*$ & 65 & $<4.5$ & $>0.39$    & 1.7$\pm$0.2     & $<16.9$ & $<47.9$  &    $<0.39$ & $<-0.56$    & $<3.14$ \\
& 3$^*$ & 94 &$<4.5$& $>0.37$   & 2.2$\pm$0.3     & $<16.9$ & $<48.3$  &     $<0.28$ & $<-0.22$   & $<3.89$ \\
& 4$^*$ & 102 & $<1.8$ & $>0.76$   & 2.4$\pm$0.5     & $<17.0$ & $<48.9$ &     $<0.34$ & $<-0.36$    & $<3.78$ \\
& 5 & 132 & 0.6$\pm$0.3 & 1.3$\pm$0.4    & 2.2$\pm$0.1     & 17.1$\pm$0.2 & 48.9$\pm$    0.3  &     0.27$\pm$0.13 & -1.8$\pm$1.3     & 1.8$\pm$ 1.5 \\
Spherical& 6 & 183 & 1.2$\pm$0.1 & 1.20$\pm$0.05    & 2.7$\pm$0.2     & 16.80$\pm$0.05 & 49.0$\pm$    0.1  &     0.12$\pm$0.01 & -1.3$\pm$0.3     & 2.9$\pm$ 0.4 \\
$f_A=1$& 7 & 254 & 1.1$\pm$0.1 & 1.7$\pm$0.1    & 3.0$\pm$0.2     & 16.93$\pm$0.05 & 49.5$\pm$    0.05  &     0.12$\pm$0.01 & -1.3$\pm$0.3     & 3.0$\pm$ 0.4 \\
$f_V=4/3$& 8 & 296 & 1.1$\pm$0.2 & 2.1$\pm$0.1    & 2.7$\pm$0.2     & 16.96$\pm$0.07 & 49.34$\pm$    0.08  &     0.11$\pm$0.02 & -1.4$\pm$0.5     & 2.7$\pm$ 0.6 \\
$\Gamma=1$& 9 & 350 & 1.1$\pm$0.2 & 2.3$\pm$0.2    & 2.4$\pm$0.2     & 16.93$\pm$0.06 & 49.06$\pm$    0.07  &     0.09$\pm$0.01 & -1.5$\pm$0.4     & 2.5$\pm$ 0.5 \\
& 10 & 459 & 1.2$\pm$0.1 & 4.4$\pm$0.3    & 3.7$\pm$0.2     & 17.14$\pm$0.04 & 50.51$\pm$    0.05  &     0.11$\pm$0.01 & -1.1$\pm$0.2     & 3.5$\pm$ 0.4 \\
& 11 & 666 & 0.9$\pm$0.1 & 4.6$\pm$0.2    & 2.7$\pm$0.1     & 17.18$\pm$0.05 & 49.80$\pm$    0.05  &     0.08$\pm$0.01 & -1.5$\pm$0.3     & 2.5$\pm$ 0.4 \\
& 12 & 749 & 0.8$\pm$0.2 & 6.5$\pm$1.1    & 3.3$\pm$0.1     & 17.4$\pm$0.1 & 50.6$\pm$    0.1  &     0.11$\pm$0.03 & -1.4$\pm$0.7     & 2.8$\pm$ 1.0 \\
& 13 & 849 & 1.1$\pm$0.2 & 2.7$\pm$0.6    & 3.1$\pm$0.3     & 17.0$\pm$0.1 & 49.8$\pm$    0.1  &     0.05$\pm$0.01 & -1.3$\pm$0.6     & 3.0$\pm$ 0.9 \\

\hline

& 2$^*$ & 65 & $<4.5$ & $>0.39$    & 1.7$\pm$0.2     & $<17.2$ & $<48.3$ &     $<0.75$ & $<-0.87$    & $<2.5$\\
& 3$^*$ & 94 &$<4.5$& $>0.37$   & 2.2$\pm$0.3     & $<17.2$ & $<48.8$  &     $<0.57$& $<-0.51$     & $<3.3$ \\
& 4$^*$ & 102 & $<1.8$ & $>0.76$   & 2.4$\pm$0.5  & $<17.4$ & $<49.4$   &    $<0.67$ & $<-0.64$    & $<3.2$ \\
& 5$^*$ & 132 & 0.6$\pm$0.3 & 1.3$\pm$0.4    & 2.2$\pm$0.1     & 17.5$\pm$0.2 & 49.4$\pm$    0.3  &     0.46$\pm$0.22 & -2.1$\pm$1.3     & 1.2$\pm$ 1.5 \\
Conical& 6 & 183 & 1.2$\pm$0.1 & 1.20$\pm$0.05    & 2.7$\pm$0.2     & 17.17$\pm$0.05 & 49.6$\pm$    0.1  &     0.24$\pm$0.03 & -1.6$\pm$0.3     & 2.4$\pm$ 0.4 \\
$f_A=0.13$& 7 & 254 & 1.1$\pm$0.1 & 1.7$\pm$0.1    & 3.0$\pm$0.2     & 17.31$\pm$0.05 & 50.0$\pm$    0.1  &     0.24$\pm$0.03 & -1.5$\pm$0.3     & 2.5$\pm$ 0.4 \\
$f_V=1.15$& 8 & 296 & 1.1$\pm$0.2 & 2.1$\pm$0.1    & 2.7$\pm$0.2     & 17.33$\pm$0.07 & 49.84$\pm$    0.08  &     0.22$\pm$0.04 & -1.7$\pm$0.5     & 2.2$\pm$ 0.6 \\
$\Gamma=1$& 9 & 350 & 1.1$\pm$0.2 & 2.3$\pm$0.2    & 2.4$\pm$0.2     & 17.30$\pm$0.06 & 49.54$\pm$    0.07  &     0.18$\pm$0.03 & -1.8$\pm$0.4     & 1.9$\pm$ 0.5 \\
& 10 & 459 & 1.2$\pm$0.1 & 4.4$\pm$0.3    & 3.7$\pm$0.2     & 17.52$\pm$0.04 & 51.09$\pm$    0.05  &     0.22$\pm$0.02 & -1.3$\pm$0.2     & 3.0$\pm$ 0.4 \\
& 11 & 666 & 0.9$\pm$0.1 & 4.6$\pm$0.2    & 2.7$\pm$0.1     & 17.55$\pm$0.05 & 50.3$\pm$    0.1  &     0.17$\pm$0.02 & -1.8$\pm$0.3     & 2.0$\pm$ 0.4 \\
& 12 & 749 & 0.8$\pm$0.2 & 6.5$\pm$1.1    & 3.3$\pm$0.1     & 17.7$\pm$0.1 & 51.1$\pm$    0.1  &     0.23$\pm$0.06 & -1.7$\pm$0.7     & 2.3$\pm$ 1.0 \\
& 13 & 849 & 1.1$\pm$0.2 & 2.7$\pm$0.6    & 3.1$\pm$0.3     & 17.4$\pm$0.1 & 50.3$\pm$    0.1  &     0.11$\pm$0.02 & -1.5$\pm$0.6     & 2.5$\pm$ 0.9 \\
		\hline
	\end{tabular}\\
	\textit{Note:}All times are reported with reference to $t_0$, MJD 58522.\\ $^*$ The peak of the synchrotron spectrum is not well-constrained by the radio observations. 
\end{table*}

A simple linear fit (assuming constant velocity) to the predicted radius gives an outflow launch date of MJD 58435$\pm$10 (2018 Nov 13, concial) or MJD 58432$\pm$10 (2018 Nov 10, spherical), approximately 120\,d before the first radio detection on MJD 58551 (2019 Mar 09). The optical/UV flare was observed to begin on MJD 58528 (2019 Feb 14),  $\sim90$\,d after the estimated outflow launch date, and peaked on approximately MJD 58560 \citep[2019 Mar 18][]{Liu2019,Hinkle2021}. From a power-law fit of the optical rise observed by ASASSN, \citet{Hinkle2021} inferred that the time of first light for the event was MJD 58522 (2019 Feb 8), indicating the radio outflow was likely launched later than MJD 58433 with an initial velocity higher than 0.1\,$c$. Thus in this work we assume an outflow launch date of MJD 58522; coincident with the optical time of first light.

\subsection{Expected future evolution of the outflow in the Sedov-Taylor decay phase}

The predicted evolution with time of the outflow's velocity (Figure \ref{fig:blastwavemodel}) is consistent with either an outflow expanding with constant velocity until the last three epochs (post-radio peak), or a gradually decelerating outflow. In the case that the outflow had approximately constant velocity of $\approx0.1$\,$c$ until the last three epochs, we suggest that this could be indicative of an outflow that was ``coasting" until the peak radio flux was reached ($t\approx650$\,d), and is now decelerating as the flux decays. However, the data could be equally well explained by a model in which the outflow consistently decelerates over the course of the observations. Under the assumption that the outflow did exhibit a coasting phase, the outflow sweeps up material from the circumnuclear medium (CNM), increasing the energy released in the emitting region as the outflow impacts the CNM \citep[e.g.][]{Generozov2017}. The onset of deceleration could indicate the outflow is entering the Sedov-Taylor phase of its evolution, at which time the outflow has reached peak energy/flux emission in the free expansion phase by sweeping up mass from the CNM, and begins to decelerate with constant energy as the blastwave approaches spherical symmetry \citep[e.g.][]{Sironi2013}. In Figure \ref{fig:blastwavemodel} there is some evidence for deceleration of the outflow in the last three epochs of observations from 660\,d post-disruption, corresponding to the epochs post-radio luminosity peak, with the deceleration most evident in the final epoch. We note, however,  that an alternate velocity evolution consisting of a power-law evolution in radius and velocity with time fits the radius measurements equally well. 

\begin{figure}
    \centering
    \includegraphics[width=\columnwidth]{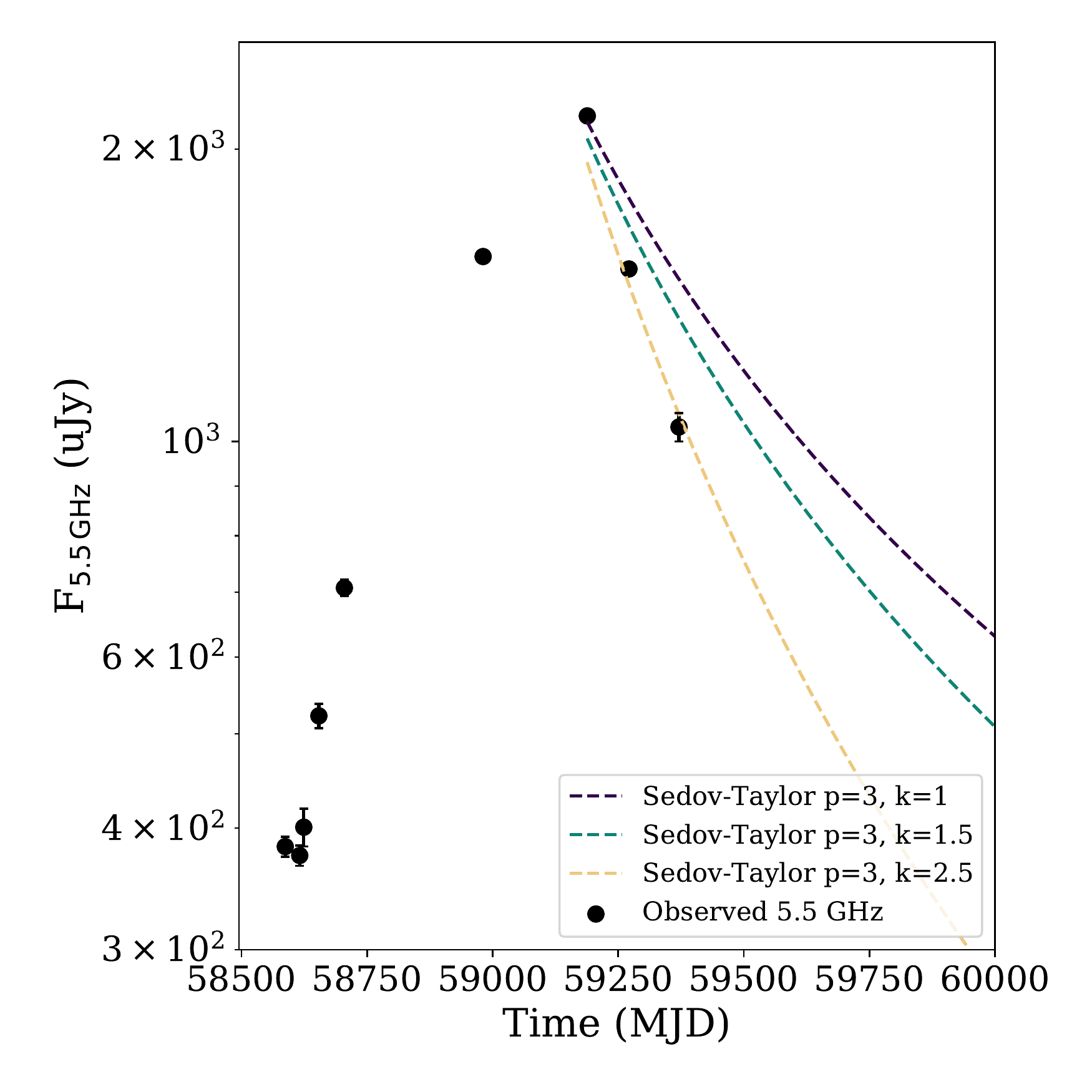}
    \caption{The observed 5.5\,GHz lightcurve of AT2019azh (black) and the predicted decay rates for a Sedov-Taylor solution with different CNM density stratifications ($n\propto r^{-k}$) for $p=3$ (dashed lines). The current decay-rate at 5.5\,GHz indicates a steeper CNM density stratification is preferred, with $k\approx2.5$.}
    \label{fig:STlightcurve}
\end{figure}

Under the assumption that the outflow velocity only began decelerating after the lightcurve peaked, the Sedov-Taylor phase enables predictions about the rate of decay of the emission for observing frequencies much above the self-absorption frequency and much below the cooling frequency for a stratified CNM density profile and the synchrotron spectrum power law index $p$ \citep{Sironi2013}. In Figure \ref{fig:STlightcurve} we show the Sedov-Taylor solution for different CNM density stratifications with $p=3$ (suitable for the final epoch of observations) for the 5.5\,GHz light curve of AT2019azh. The predicted flux evolution is calculated assuming late-time radio emission in the Sedov-Taylor phase, in which a spherical shock runs into a stratified medium with density profile $n\propto r^{-k}$ and the electrons in the shock are accelerated into a synchrotron power-law distribution \citep{Sironi2013}. The best-fit solution is for a steep CNM density gradient, with $k=2.5$, similar to the density gradients observed for ASASSN-14li and AT2019dsg (Figure \ref{fig:TDEcomp}). The current decay-rate of the radio emission at 5.5\,GHz is well-fit by the Sedov-Taylor approximation for a CNM density $n\propto^{-2.5}$ and $p=3$. 

\section{Multiwavelength modelling: accretion disk evolution}\label{sec:multiwavelength_modelling}

To assess the possibility that the radio outflow from AT2019azh was produced by accretion onto the SMBH, in this section we model the accretion disk emission based on the optical, UV, and X-ray properties of the event. The evolving disk density profiles are calculated using the method developed in \citet{Mummery2020} and \citet{Mummery2021}, to which the reader should refer for detailed information. We assume that the black hole environment is initially completely devoid of material, before feeding disk material into a ring according to the following relationship:
\begin{equation}\label{source}
\dot M_{\rm feed} \propto \delta(r-r_0) \, \left( t + \Delta t \right)^{-5/3}
\end{equation}
where $\delta$ is the Dirac delta function and $\Delta t$ ensures the feeding rate is finite as $t \rightarrow 0$, and was taken equal to one code time step. The feeding radius was taken to be $r_0 = 50R_g$, appropriate for a TDE around a low-mass black hole like AT2019azh, and the disk evolution was started at a time corresponding to the first observed optical emission of the source. This model assumes that the matter is fed into the disk at the rate at which disrupted stellar material returns to pericentre \citep[the so-called ‘fall-back’ rate $\dot M_{\rm fb}$,][]{Rees1988}. The material fed into the disk then evolves according to the equations of disk angular momentum conservation and disk mass conservation. Energy conservation then allows the disk temperature to be calculated at each radius and time. We assume that each disk radius emits like a colour-corrected blackbody (using the \citet{Done2012} colour-correction model), and ray-trace the resulting disk emission profile. We include all relevant relativistic effects, such as Doppler and gravitational redshift, and gravitational lensing. 
The evolution of the X-ray and UV light curves of a thermal TDE are therefore determined by three fitting parameters: the black hole mass $M$, the total accreted mass  $M_{\rm acc}$ (a normalisation on the source term, Eq. \ref{source}), and the viscous timescale of the evolving disk $t_{\rm visc}$. We compute $M_{\rm acc}$ in the following manner
\begin{equation}
M_{\rm acc} \equiv \int_0^\infty \dot M(r_I, t) \, {\rm d} t,
\end{equation} 
where $\dot M(r_I,t)$ is the inner-most stable circular orbit (ISCO) mass accretion rate. 

Given the simplifications applied in the modelling (the black hole spin is fixed to zero, and the observer inclination angle is fixed to $\theta_{\rm obs} = 30^\circ$) the  best fit parameter values and their associated uncertainties should be treated with some caution, although \citet{Mummery2020} found only a moderate change (factor of $\sim 1.5$) in best fit black hole mass over a wide range of black hole spins. 

Finally, we determine the best fit system parameters by simultaneously minimizing the sum of the squared differences between the model and the different UV and X-ray light curves of AT2019azh. As in previous works, we anticipate large formal values of the reduced $\chi^2$. These large values result as a consequence of short time-scale fluctuations present in the well-sampled TDE light curves, and are to be expected in any theoretical model using a smooth functional form for the time-dependence of the turbulent stress tensor. Our standard approach implicitly averages over rapid turbulent variations. Short time-scale fluctuations are likely to be highly correlated so that accurately assessing the statistical significance of the fit is not straightforward. We have therefore used $\chi^2$ minimisation as a sensible guide towards finding a best fit.

\begin{figure*} 
    \centering
    \includegraphics[width=0.45\textwidth]{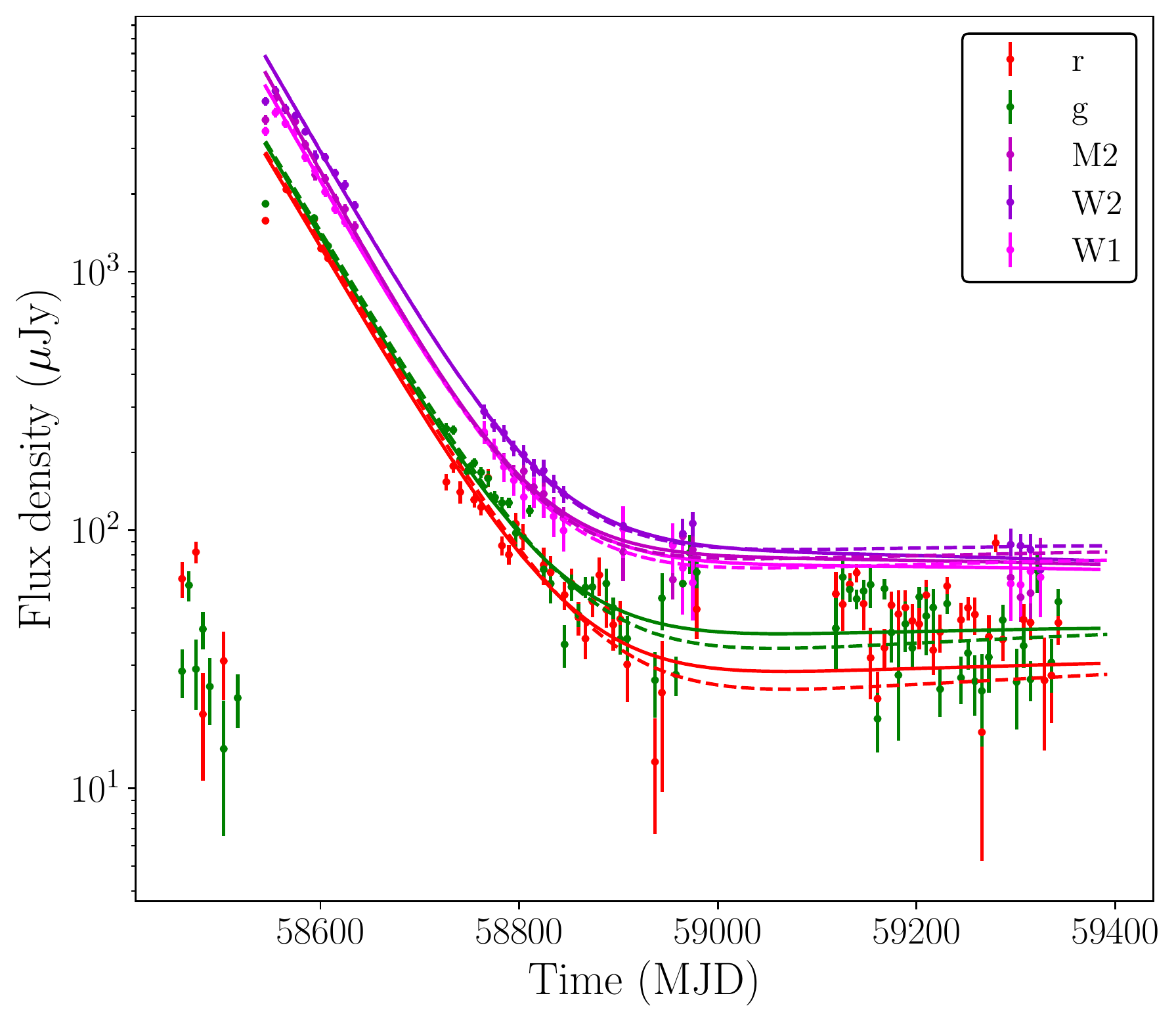}
    \includegraphics[width=0.45\textwidth]{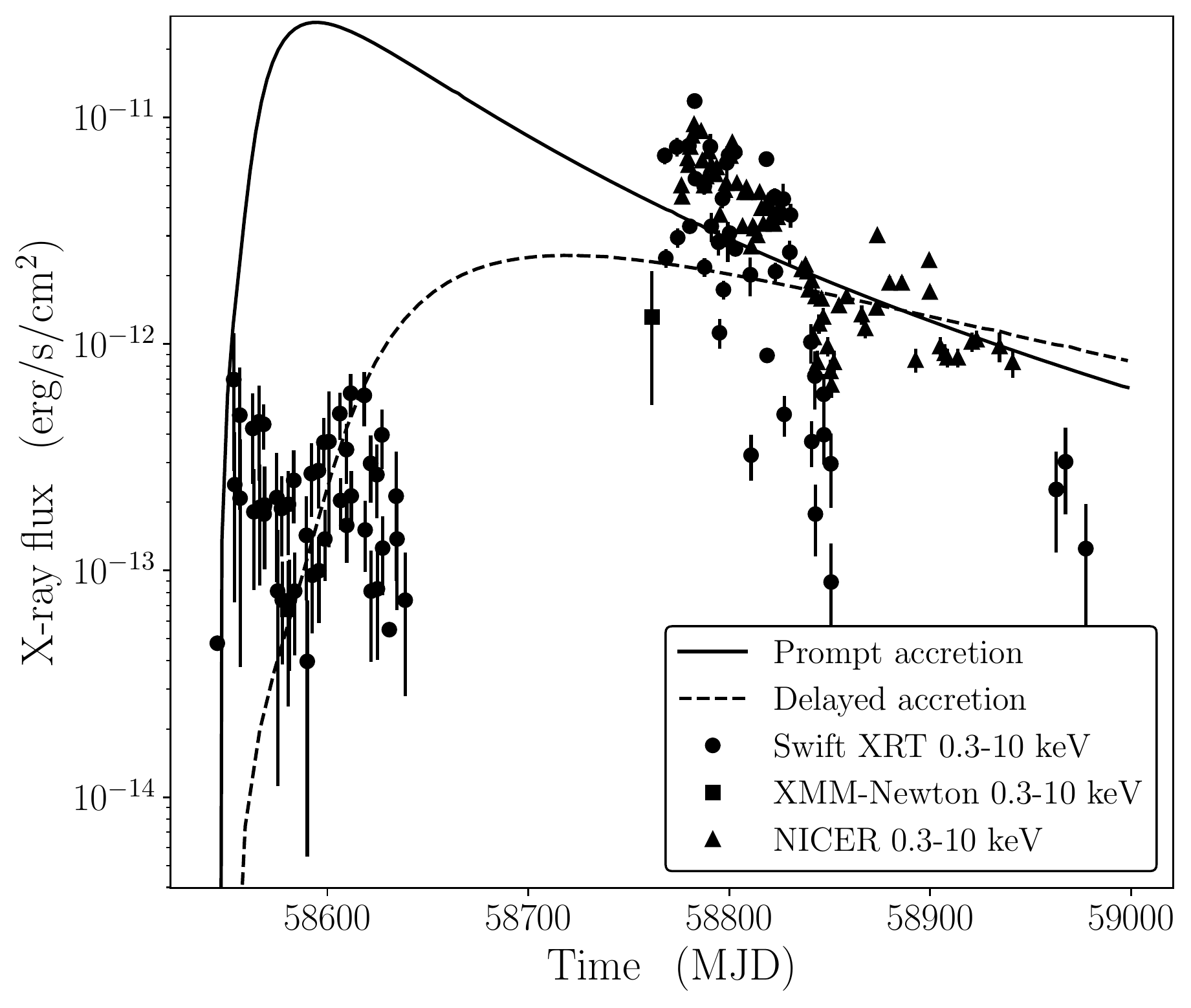}
    \includegraphics[width=0.6\textwidth]{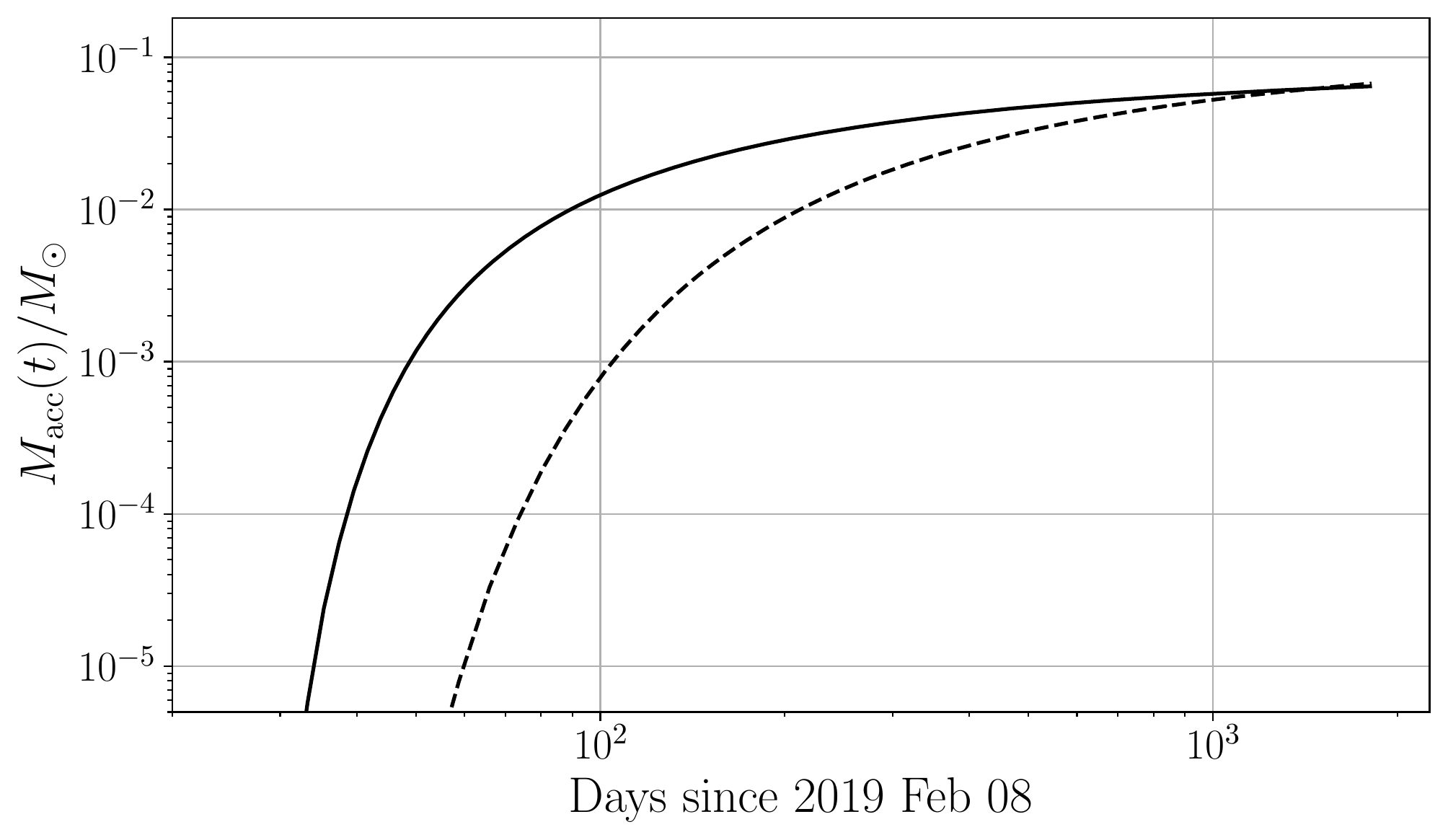}
    \caption{The observed values, shown as points with error bars, and the modelled values, shown by solid or dashed lines of the optical/UV (top left) or X-ray (top right) properties of the disk and modelled accreted mass (mass past the ISCO radius, bottom) in the TDE AT2019azh. The ZTF $r$- and $g$-band observations are shown in red and green respectively and the \textit{Swift} UVM2 filter and 2-10\,keV X-ray observations are shown in magenta and black respectively. We model two scenarios: a delayed disk-formation (dashed lines) and an early disk-formation (solid lines). The optical and UV observations are well-fit by either model, while the X-ray observations are better-fit by the delayed accretion scenario, unless in the case of significant X-ray obscuration at early times. The X-ray observations are from \citet{Hinkle2021} and only the 0.3-10\,keV data from each telescope was used in the fitting.}
    \label{fig:Andymodelling}
\end{figure*}

The observed and modelled optical, UV, and X-ray lightcurves for AT2019azh are plotted in Figure \ref{fig:Andymodelling} for two scenarios: early accretion and delayed accretion. The observed optical and UV data are well fit by both the early and delayed accretion models, with little difference between the two. The observed X-ray lightcurve is significantly better fit by the delayed accretion model at early times, whilst the late-time X-ray lightcurve is well-fit by either model. We note that in the case of significant X-ray obscuration at early times the early accretion model may also be viable.

\subsection{Different disc evolution scenarios}
The X-ray evolution of AT2019azh is, as discussed in the Introduction, somewhat atypical for a TDE. Thus, we present two models for the light curve evolution of AT2019azh: `prompt' and `delayed' accretion. 
There are two ways in which the peak X-ray luminosity of a TDE may be delayed. The X-ray luminosity of a TDE is primarily a function of the hottest temperature in the TDEs disc \citep{Mummery2020}, as the TDEs peak X-ray luminosity is only reached once the TDEs inner disc density reaches its maximum. As typical TDE feeding radii are tens to hundreds of gravitational radii, a large viscous timescale (which delays the build up of the inner disc density by increasing the length of time it takes for the disc material to propagate inwards) can suppress early TDE X-ray emission. Alternatively, if there is substantial obscuration of the inner regions of the accretion disc at early times, which then clears at larger times, this may also lead to a late-time rise in X-ray luminosity. We model both scenarios in this section.  
In the delayed accretion model we fit to the entire X-ray light curve, finding as expected a large viscous timescale $t_{\rm visc} = 220 \pm 20$ d. The accreted mass, $M_{\rm acc} = 0.1 \pm 0.02 M_\odot$, is consistent with a star of stellar mass $M_\star \simeq 0.2 M_\odot$ \citep[i.e., no missing energy problem, as in][]{Mummery2021b}. The best fitting black hole mass $M_{\rm BH} = 3.2^{+0.15}_{-0.1} \times 10^6 M_\odot$, is consistent with the galaxy scaling measurement $M_{\rm BH} < 4 \times 10^6 M_\odot$ \citep{vanVelzen2016}. The peak Eddington ratio of this model is $l_{\rm peak} = 0.22$ ($L_{\rm peak} = 8.5 \times 10^{43}$ erg/s), and the total radiated energy was $E_{\rm rad} \simeq 8 \times 10^{51}$ erg. 
In the prompt accretion model, which requires substantial early time obscuration, we find a viscous timescale $t_{\rm visc} = 65 \pm 10$ d. The accreted mass, $M_{\rm acc} = 0.07 \pm 0.01 M_\odot$, and best fitting black hole mass $M_{\rm BH} = 2.1^{+0.15}_{-0.1} \times 10^6 M_\odot$ are again consistent with their expected values. The peak Eddington ratio of this model is $l_{\rm peak} = 0.75$ ($L_{\rm peak} = 2.1 \times 10^{44}$ erg/s), and the total radiated energy was $E_{\rm rad} \simeq 5 \times 10^{51}$ erg. 
The difference between the two accretion evolution scenarios can be seen in Figure \ref{fig:Andymodelling}, which shows  the accreted mass as a function of time (plotted in the same units as Figure \ref{fig:blastwavemodel}).

\section{Discussion}\label{sec:Discussion}
Our findings reveal a likely non-relativistic outflow with constant (or gradually decreasing) velocity and continuous kinetic energy increase during the radio rise (up to 666\,d), and constant energy post-radio peak (from 666--849\,d). We infer that the outflow ranges from radii of $\sim3\times10^{16}$\,cm--$2\times10^{17}$\,cm with energies of $\sim3\times10^{47}$\,erg--$1\times10^{51}$\,erg. These energy and radii correspond to a magnetic field of $\sim0.05$\,G and ambient electron density of $\sim50$--3000\,cm$^{-3}$. The observed energy and radius increased with time until the peak radio luminosity was reached. 

The optical and X-ray observations, particularly the late-time X-ray rise, have been explained with either a delayed-accretion disk formation scenario \citep{Liu2019} or due to an increase in the X-ray emitting region \citep{Hinkle2021}. We modelled two disk emission scenarios in Section \ref{sec:multiwavelength_modelling}, and found that the observed UV/optical behaviour of the event was well-fit by either model (Figure \ref{fig:Andymodelling}). The X-ray emission is better fit by a delayed accretion scenario except in the case of significant X-ray obscuration at early times. 
%
\subsection{The unusual late time steepening of the synchrotron spectrum}

We observed statistically significant fluctuations (at a 3-$\sigma$ level) in the synchrotron energy index, $p$, of the optically-thin part of the synchrotron spectrum from $\sim450$\,d post-disruption. We found that the energy index steepened to $p=3.7\pm0.2$ at $t=460$\,d, reducing back to $p\approx2.6$ at $t=660$\,d, steepening again to $p=3.3\pm0.1$ at $t=750$\,d, and finally reducing back to $p\approx 3$ in the final epoch at $t=850$\,d post-disruption. The mean value of $p$ for the epochs without spectral steepenings is $2.7\pm0.2$ (1-$\sigma$ error, excluding the first 4 epochs where the spectra were not well constrained), indicating the spectral steepening at $t=460$\,d is significant to 3-$\sigma$, and the steepening at $t=750$\,d is significant to 2-$\sigma$. After detailed investigation (see Appendix \ref{sec:appendix1}), we found that these fluctuations are not due to calibration issues with the data or inconsistent flux density scaling between epochs as there is no such systematic difference in three background sources that we examined in the field of view for each epoch (excepting the first minor steepening at 200\,d, which we conclude is not statistically significant, see Appendix \ref{sec:appendix1}). 

Fluctuations in the energy index have not been observed in the radio emission from thermal TDEs to date, and are difficult to explain in the current (single-zone) synchrotron emission model. Usually the steepening of a synchrotron spectrum can be attributed to adiabatic cooling of the electrons, and would indicate the detection of a cooling break in the spectrum \citep{Granot2002}. However, the adiabatic cooling timescales are too long to explain the fluctuations on timescales of $\sim$months that we observed in this event, and we find no evidence for the presence of a cooling break in any of the spectra (see Appendix \ref{sec:appendix2}). 

We propose that the energy index fluctuations could be due to a spherical or collimated outflow encountering an inhomogenous CNM, or fluctuations in the energy injection rate of a collimated jet-like outflow. In the spherical outflow scenario, different populations of electrons from different regions of the outflow might encounter inhomogenous clumps of the CNM, changing the emitting properties of different populations of electrons, each with their own synchrotron spectrum. Smaller populations will fade quickly, contributing less to the total radio emission, and allowing the flux to fall at higher frequencies. The synchrotron spectra we observe at each epoch is the sum of the emission from these different populations of electrons. If the fluctuations in $p$ are due to the changes in the shock acceleration efficiency, we expect more fluctuations in the radio lightcurve at 9\,GHz than at 5\,GHz, as indeed is seen in Figure \ref{fig:lightcurve}.

\subsection{The outflow mechanism}\label{sec:outflowmodels}
With the addition of the radio observations to the multiwavelength data we are able to obtain a more robust picture of the event and how the different types of emission may have been produced.  The empirical properties of the outflow in AT2019azh obtained partly from the radio observations are key to modelling and understanding the mechanism that produced the outflow. What is crucial to understanding the event is that the radio outflow was first observed early, around the time of disruption, and well before the X-ray emission brightened, in contrast to the suggestion that delayed radio emission is common in TDEs \citep{Horesh2021b}. Furthermore, the energy index fluctuations observed in the radio emission place strong constraints on the geometry of the outflow; a spherical homogeneous outflow cannot produce the observed fluctuations. Below we discuss different scenarios that could explain the observed properties of the outflow in AT2019azh. 

\subsubsection{Accretion-driven wind outflow}
Accretion onto an SMBH can produce winds and outflows that would be observable in the radio as they travel through the CNM at velocities $\sim0.01-0.1\,c$ \citep[e.g.][]{Mohan2021,Strubbe2009,Tchekhovskoy2014}. A popular model to explain non-relativistic outflows from TDEs is a spherical wind driven by accretion onto the SMBH \citep[e.g.][]{Alexander2016,Cendes2021}. In this scenario, the radio outflow should appear at approximately the time that high accretion luminosities are observed at X-ray wavelengths, provided there is no obscuration of the X-ray emission. A wind outflow would have approximately spherical geometry \citep[e.g.][]{Mohan2021}, and produce radio emission from a forward shock from the non-relativistic outflow expanding into the CNM driven by the gas accretion onto the SMBH. 

The early radio emission for AT2019azh is difficult to explain as an accretion wind-induced outflow in the delayed accretion scenario, due to the lack of bright X-ray emission indicative of significant accretion (requiring $F_X\sim10^{11}$\,erg/s/cm$^2$) and the lack of any X-ray/optical correlation, unless there was strong obscuration of the X-ray emission. In Section \ref{sec:multiwavelength_modelling}, we found that the optical, UV, and X-ray observations are well-fit by either delayed accretion or early accretion with significant X-ray obscuration (Figure \ref{fig:Andymodelling}). In the delayed accretion scenario, the radio outflow could not be produced by an accretion-driven wind due to the lack of significant amounts of material reaching the black hole to be ejected into the outflow at early times, when the first radio emission was observed.

\subsubsection{Sub-relativistic jet}
In the scenario of a mildly relativistic or sub-relativistic jet, a collimated outflow is produced by accretion onto the SMBH and the radio emission could be produced by either a forward shock that the jet drives into the surrounding medium, or internally through shocks inside the jet \citep[e.g.][]{vanVelzen2016}, both of which would produce synchrotron emission. A sub-relativistic or mildly relativistic jet was proposed initially for ASASSN-14li \citep{vanVelzen2016} and AT2019dsg \citep{Stein2021}. The main argument against a jet-like outflow relies on the geometric factors, and the level of collimation required to obtain a self-consistent solution for the outflow properties. 

We deduce that a relativistic jet explanation for the radio properties of AT2019azh is not supported by the data. Similar to the argument against a relativistic jet provided in \citet{Alexander2016}, if we introduce an additional parameter, the bulk Lorentz factor ($\Gamma$) to the synchrotron equipartition model outlined in Section \ref{sec:Modeling} \citep{BarniolDuran2013}, in order to obtain a self-consistent result where $\Gamma\gtrsim2$ (i.e. a relativistic outflow) requires $f_A\lesssim 0.01$, i.e. a jet with opening angle of $\lesssim0.1\deg$. Such a small opening angle is not possible for SMBH outflows \citep[e.g.][]{Jorstad2005} and rules out the possibility of a relativistic jet for the outflow from AT2019azh. 

The late-time evolution of a sub-relativistic jet and a mildly relativistic spherical outflow appear very similar at radio frequencies \citep{Nakar2011}, however, early on an initially on-axis relativistic jet that decelerates to non-relativistic velocities would appear much more energetic (with energies comparable to Sw J1644+57 $\sim10^{52}$\,erg). The luminosity we observed for AT2019azh ($L\sim10^{38}$\,erg/s) disfavours the possibility of an initially relativistic on-axis jet for the early radio emission. With observations of this event spanning the peak of the radio lightcurve, we can also deduce that the outflow is likely non-relativistic due to the observed behaviour of the lightcurve as the outflow transitioned from freely expanding to decelerating. The Doppler factors are no longer important when the outflow begins decelerating \citep[e.g.][]{Sironi2013} so the radius constraint obtained is the true radius of the outflow. If the radio emission was produced by an off-axis relativistic jet, at the time of deceleration it would be emitting isotropically and we would expect to see a flux increase, which is not observed in the radio lightcurve. Under the assumption that the outflow transitioned into the sub-relativistic Sedov-Taylor decay phase after the peak of the radio lightcurve (Figure \ref{fig:STlightcurve}), the inferred radius at the time of transition yields an average speed of the outflow that is significantly less than the speed of light. This would further confirm the sub-relativistic nature of this event, in contrast to the assumed relativistic event Arp 299-B AT1, which was found to initially move at relativistic speeds for the first $\sim760$\,d \citep{Mattila2018}. Alternatively, in the scenario in which the outflow was constantly decelerating over the course of the radio observations, we cannot rule out initially relativistic speeds of the outflow prior to the first radio detection.  

A sub-relativistic jet, with $\Gamma\approx1$, would not require such extreme collimation of the emission. A sub-relativistic jet may present similarly to our conical geometry model (Table \ref{tab:outflowproperties} and Figure \ref{fig:blastwavemodel}). Such a jet would have slightly larger radii, higher velocity, increased energy and require a lower CNM density to self-consistently explain the observed properties of the emission, than would a spherical outflow. A collimated outflow may also explain the energy index fluctuations in the case of an inhomogenous CNM. As a sub-relativistic jet travels through the CNM, it would sweep up material, slowing down and causing the jet to spread laterally \citep[e.g.][]{vanVelzen2016}. In this scenario the emission from the jet is more isotropic due to the Doppler factor close to unity, and a narrow viewing angle is not necessary in order to observe the radio emitting region. A sub-relativistic jet driven by accretion may have continued energy injection until the central engine switches off \citep[e.g.][]{Mohan2021}, which could explain the continuous energy increase observed for AT2019azh (Figure \ref{fig:blastwavemodel}). Furthermore, the material ejected from close to the black hole could easily reach velocities and energies as high as we predict for AT2019azh due to energy conservation and the angular momentum available at the inner orbits of the SMBH. However, through our disk modelling (Section \ref{sec:multiwavelength_modelling}), we infer that the accretion rate never exceeds 0.2 times the Eddington rate in the delayed accretion scenario, and 0.7 in the early accretion scenario. For the observed radio emission to be explained by an accretion-driven outflow it would require inefficient accretion onto the SMBH, which has not been confirmed in observations of SMBHs.
Thus we deduce that a mildly collimated sub-relativistic jet may explain the observed properties of the outflow in AT2019azh, under the condition that the accretion disk formation was not delayed and there was significant X-ray obscuration at early times. 

\subsubsection{Collision induced outflow}
\citet{Lu2019} modelled the self-intersection of tidal debris streams during a TDE and deduced that during stream-stream collisions of the tidal debris, a significant amount of gas will become unbound and ejected as a collision induced outflow (CIO). This kind of outflow would have kinetic energies between $10^{50}$--$10^{52}$\,erg and velocities between 0.01--0.1\,$c$, similar to the properties we infer for the outflow of AT2019azh (Figure \ref{fig:blastwavemodel}).

Due to the lack of evidence at early times of significant accretion that could produce an outflow from inefficient accretion onto the SMBH (unless in the case of significant X-ray obscuration), the CIO model is quite promising to explain the radio emission from AT2019azh. A CIO will be launched when the streams intersect, an event that precedes the start of accretion onto the black hole, which could well explain the early radio detection. \citet{Liu2019} proposed that a collision induced outflow model could explain the UV/optical and X-ray emission from AT2019azh at early times as the debris is becoming circularised; our radio detection pre-optical peak supports this theory. 

In the case of a CIO, the outflow would be produced by a prompt injection of energy during the circularisation, and the outflow emission would evolve over time as the spherical ejecta is slowly shocked by the CNM and sweeps up material. In order to reach velocities as high as $0.1\,c$, the pericenter of the destroyed star would need to be within 10--15\,$R_g$ of the SMBH (depending on the black hole spin), otherwise the CIO would not be strong enough to reach the observed velocities and energies of AT2019azh.  A CIO outflow is well-described by a ``coasting", free-expansion and a Sedov-Taylor decay phase once the peak luminosity is reached \citep{Lu2019}, in contrast to a jet-like outflow which could begin with an expansion phase during which the outflow is powered by the jet-engine, then a Sedov-Taylor decay phase when the jet switches off. The deceleration radius and transition from the free-expansion to Sedov-Taylor phase corresponds to the time at which the lightcurve peaks, and is characterised by $E_{\rm{k}}=(1/2)N(r_{\rm{dec}}m_{\rm{p}}v_0^2$ \citep[e.g.][]{Lu2019}, i.e. the deceleration is caused by the outflow interacting with the CNM. 

\subsubsection{Unbound debris stream}
When a star is destroyed by a SMBH, approximately half of the stellar debris will be captured by the gravitational well of the black hole to be accreted, whilst the remaining half of the star is unbound, and may be ejected from the system with high escape velocities ($v>10^4$\,km\,s$^{-1}$) \citep{Rees1988}. This unbound debris will interact with the circumnuclear medium, emitting synchrotron radiation in the bow shock that forms along the leading edge of the debris stream \citep[e.g.][]{Krolik2016}. The earliest emitting region will correspond to the fastest unbound debris, expanding at velocities of $\sim0.05\,c$ \citep{Krolik2016}. Over time, the bulk of the unbound debris will decelerate and eventually become visible, adding to the emitting region of the outflow. The unbound material would be confined to a very small solid angle \citep{Guillochon2014,Kochanek1994,Coughlin2016}, which is often used as a justification to rule-out radio emitting outflows being produced by the unbound debris stream \citep[e.g.][for ASASSN-14li]{Alexander2016}. 

The predicted mass in the outflow for the radio-emitting region of AT2019azh is significantly less than expected for the entire unbound debris (Figure \ref{fig:blastwavemodel}, assuming a $\sim1\,M_{\odot}$ star was destroyed and approximately half of the stellar debris is ejected in the unbound debris stream \citep[e.g.][]{Rees1988}). However, only the debris with the fastest escape velocities would be visible early on, corresponding to a very small fraction of the unbound debris. If the outflow in AT2019azh were produced by the unbound debris we may expect to see the radio evolving at later times as the slower debris catches up. However, in the unbound debris model, the outflow should continue to expand at a constant velocity without slowing down until the ejecta has swept up a mass comparable to its own \citep{Krolik2016}. In Figure \ref{fig:blastwavemodel} we find weak evidence of fluctuations in the velocity of the outflow, and some downwards trend between $t=250-850$\,d. The velocity of the outflow we infer even in the spherical geometry model ($\gtrsim0.1$\,c) is higher than expected in models of the unbound debris ($\approx0.05\,c$) and the inferred energy of the outflow ($\gtrsim10^{48}$\,erg) is also greater than expected for the unbound debris stream ($\sim10^{47}$\,erg) \citep[e.g.][]{Krolik2016}. An unbound debris stream outflow would require slightly higher collimation of the emission than we consider in the conical model ($f_A\approx0.2$) in Section \ref{sec:Modeling}, which would only increase the predicted energy and velocities. \\

We thus conclude that the multiwavelength emission of the outflow from AT2019azh could be explained self-consistently by either a collision induced outflow or, less likely, an accretion-driven wind or sub-relativistic jet. The equipartition analysis in Section \ref{sec:outflowmodels} provides a robust estimate of the size of the emitting region, and thus its velocity, but it does not enable strong discrimination between the energy source of the emission, and thus the driving source of the outflow. The early radio emission combined with the low X-ray emission and lack of early optical/X-ray correlation points towards the radio outflow not being produced through accretion onto the SMBH, unless there was significant X-ray obscuration. Our disk modelling of the multiwavelength observations in Section \ref{sec:multiwavelength_modelling} indicates that the optical/UV and X-ray emission is well-fit by either an immediate accretion scenario or a delayed accretion scenario with X-ray obscuration. The observed fluctuations of energy and $p$ could be explained by an inhomogenous CNM, in which different populations of electrons in the outflow are emitting and being observed at different times.

A more energetic outflow is possible in a disk-driven outflow than a CIO as the ejected material can be ejected with larger energy, translating to a faster, more energetic outflow, but both scenarios could reach the energies and velocities we deduce for the outflow in AT2019azh. Further, detailed modelling of the multiwavelength properties of this event, building on the detailed optical and X-ray analysis presented in \citet{Hinkle2021}, are necessary to truly gain a deep understanding of how to reconcile the X-ray, optical, and radio observations in a self-consistent way. 

\subsection{AT2019azh in the context of other TDEs}
\begin{figure*}
    \centering
    \includegraphics[width=\columnwidth]{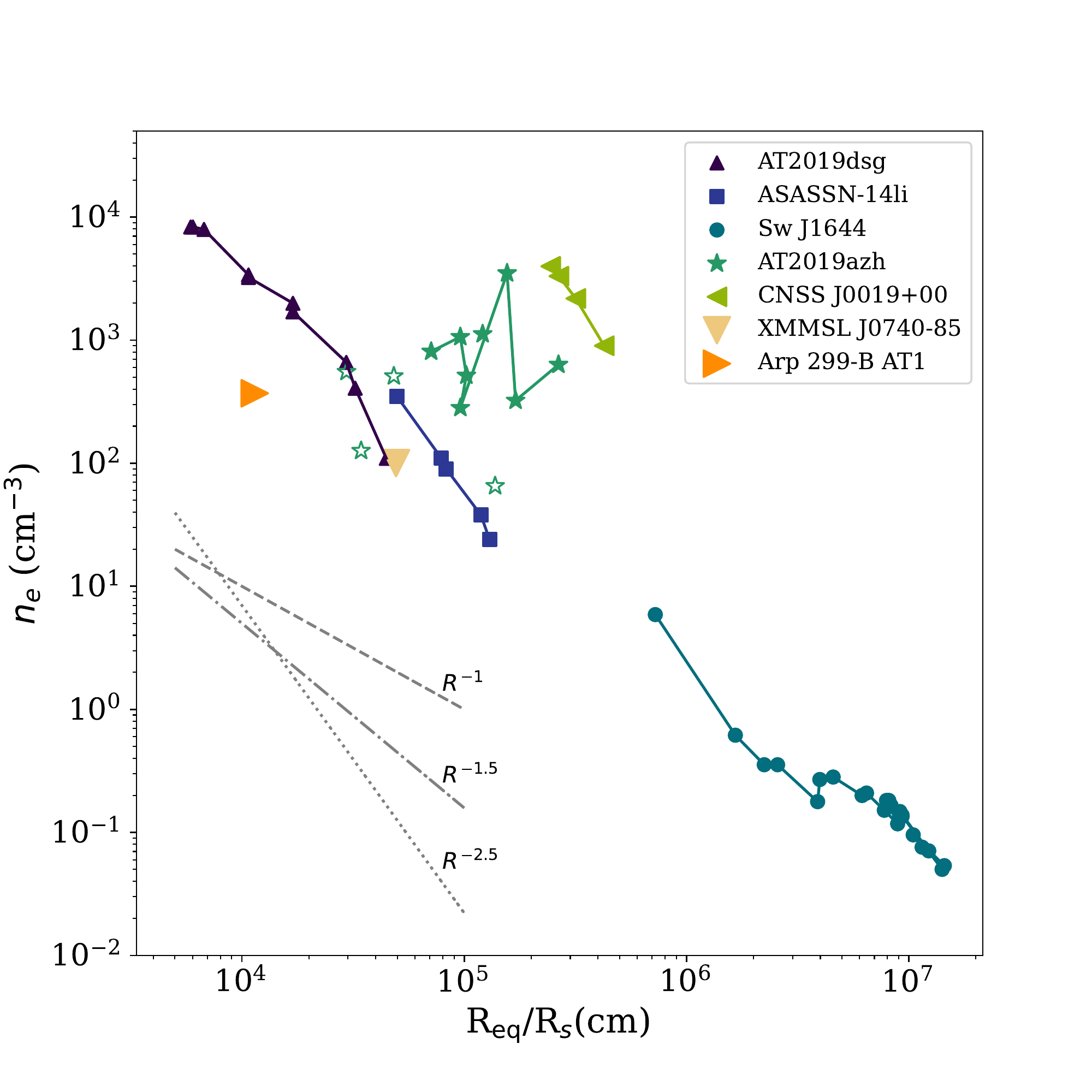}
    \includegraphics[width=\columnwidth]{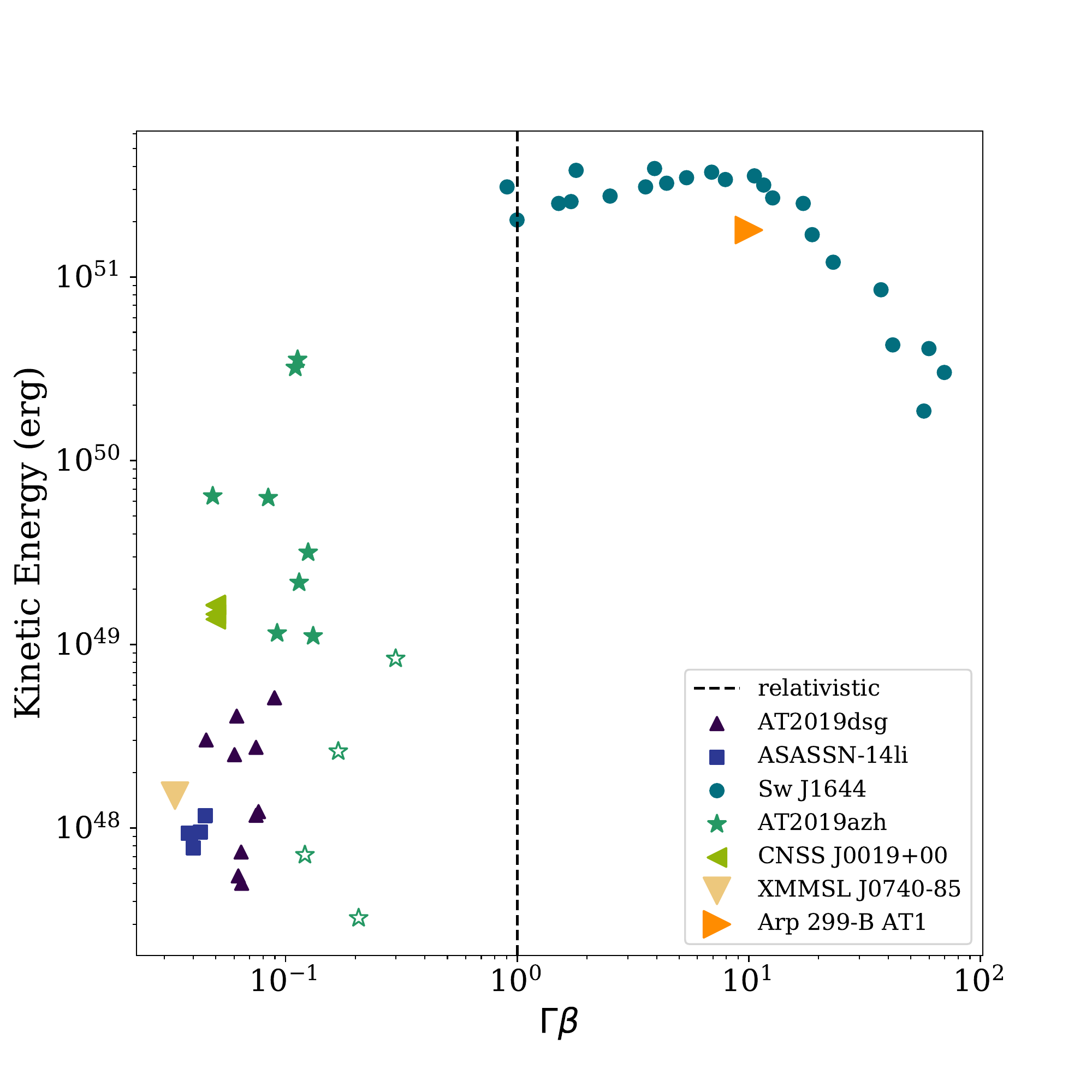}
    \caption{\textit{Left:} The scaled radius and ambient density inferred from the radio emission of known TDEs. \textit{Right:} The energy and velocity inferred from the radio emission of the known radio TDEs. AT2019azh is shown in green stars and open stars indicate the epochs where the synchrotron spectra were not well-constrained. TDE data and assumed SMBH masses are from \citet{Cendes2021,Stein2021} (AT2019dsg, $M_{\mathrm{BH}}=5\times10^6$\,$M_{\mathrm{\odot}}$), \citet{Alexander2016} (ASASSN-14li, $M_{\mathrm{BH}}=1\times10^6$\,$M_{\mathrm{\odot}}$, \citet{Eftekhari2018} (Sw J1644+57, $M_{\mathrm{BH}}=1\times10^6$\,$M_{\mathrm{\odot}}$), \citet{Anderson2020} (CNSS J0019+00, $M_{\mathrm{BH}}=1\times10^6$\,$M_{\mathrm{\odot}}$), \citet{Mattila2018} (Arp 299-B AT1, $M_{\mathrm{BH}}=2\times10^7$\,$M_{\mathrm{\odot}}$),  and \citet{Alexander2017} (XMMSL1 J0740-85, $M_{\mathrm{BH}}=3.5\times10^6$\,$M_{\mathrm{\odot}}$). For AT2019azh we assume $M_{\mathrm{BH}}=3\times10^6$\,$M_{\mathrm{\odot}}$. $R_s$ is the Schwarzschild radius of the black hole and R$_{\mathrm{eq}}$ is the predicted equipartition radius of the outflow.}
    \label{fig:TDEcomp}
\end{figure*}

On comparison of the outflow properties of AT2019azh with other TDEs with spectrally-resolved radio observations, we find AT2019azh fits well into the population of thermal TDEs (Figure \ref{fig:TDEcomp}). There is only one other thermal TDE (AT2019dsg) with multi-frequency radio coverage at early times ($t<100$\,d post-distribution). AT2019dsg showed an order of magnitude increase in energy early on, in contrast to the slow rise in radio that we observed for AT2019azh (Figure \ref{fig:lightcurve}).  

We find the ambient density is approximately proportional to $R^{-2.5}$ for most TDEs, whilst AT2019azh shows significant variation and could be described by $n\propto R^{-1}$ to $R^{-2.5}$ (Figure \ref{fig:TDEcomp}). Figure \ref{fig:TDEcomp} indicates that for higher ambient densities the radio emission is brighter and peaks earlier \citep{Lu2019}, as indeed is the case for the light curve of AT2019dsg compared to AT2019azh (Figure \ref{fig:lightcurve}).   

Inhomogeneity of the CNM density for AT2019azh compared to that of other TDEs could possibly explain the fluctuations in energy index we observed that was not seen in the other early-time TDE observations, however other studies of thermal TDE radio spectra do not comprehensively assess the possibility of variations in $p$. In Figure \ref{fig:TDEcomp} there is some evidence that the ambient density around AT2019azh is more inhomogenous than the other thermal TDEs based on the inferred spectral properties. The kinetic energy and velocity we infer for AT2019azh are somewhat larger than those for other thermal events, but does not reach the energies (or velocities) observed for relativistic events.

The late-time X-ray brightening of AT2019azh resembles the behaviour observed in the TDEs ASASSN-15oi and OGLE16aaa \citep{Horesh2021,Kajava2020}. However, in the case of ASASSN-15oi, there was no radio emission detected early on, in stark contrast to the pre-optical peak radio detection of AT2019azh. Thus, the argument for delayed accretion, while likely for ASASSN-15oi, is difficult to justify for AT2019azh if the outflow was accretion disk-driven.

The thermal TDE ASASSN-15oi also exhibited a change in energy index of the optically-thin part of the synchrotron spectrum at late times \citep{Horesh2021}. 
ASASSN-15oi exhibited some lower-level radio activity, which faded with time, $\sim100$\,d post-disruption, and a drastic increase in radio flux at $\sim600$\,d post-disruption \citep{Horesh2021}. Interestingly, the energy index of the optically thin synchrotron emission became much flatter in the late-time flare compared to the earlier emission. The early emission exhibited a standard energy index of $p=2-3$, whereas for the later flare it was much flatter at $p=0.2$ \citep{Horesh2021}. \citet{Horesh2021} deduce that the standard spherical outflow model from a super-Eddington wind cannot explain both the delayed onset of radio emission and late-time flare from this event, which latter would require an outflow to be launched at late-times, and possibly into an inhomogenous circumnuclear medium. They also propose that the emission could be explained by a transition in accretion state onto the SMBH at late-times. However, for AT2019azh we observed the opposite behaviour in $p$; a steepening of the energy index at later times, followed by fluctuations over the next months. 

The late-time radio emission of AT2019azh behaves similarly to that of the TDE AT2019dsg \citep{Stein2021}, which is explained by \citet{Cendes2021} to likely be driven by a spherical outflow from a super-Eddington wind. However, the outflow properties \citet{Cendes2021} determined for AT2019dsg could equally-well be explained by the CIO model. \citet{Stein2021} and \citet{Mohan2021} suggest that the initially increasing energy and flux density observed from AT2019dsg (as in AT2019azh) is produced by a constant energy injection at the source of the outflow, which later switches off and causes the radio emission to fade. At this point the outflow naturally decelerates due to the shutting off of the central engine, rather than due to interactions with the CNM. This could be the case for AT2019azh in the sub-relativistic jet outflow scenario, however, the increasing energy can also be explained by the outflow sweeping up CNM material at a constant velocity \citep{Sironi2013}.

\section{Summary}\label{sec:Conclusion}
We followed the radio evolution of the tidal disruption event AT2019azh for 850 days post-disruption. The radio emission rose slowly to a peak over 650\,d, and is now decaying following the expected decay rate for the Sedov-Taylor solution. We modelled the radio emission as a spherical (or conical), non-relativistic outflow and infer energies of $3.5\times10^{47}$--$2.7\times10^{50}$\,erg ($9\times10^{47}$--$1\times10^{51}$\,erg), radii of $2\times10^{16}$--$2\times10^{17}$\,cm ($5\times10^{16}$--$5\times10^{17}$\,cm), and circumnuclear density of 70--1100\,cm$^{-3}$ (15--1000\,cm$^{-3}$), for a velocity of $\approx0.1\,c$. We detected radio emission approximately 10\,d before the optical peak, which is the first radio detection of a thermal TDE at such early times, and well before the X-ray emission indicated any signs of significant accretion. This early time radio detection is in contrast to the suggestion that late-time radio flares or delayed radio emission is common in TDEs \citep{Horesh2021b}. Such an early radio detection could rule out the accretion-driven wind outflow model at early times, but through detailed disk modelling we found that both a scenario with delayed accretion or one with early accretion but significant X-ray obscuration could explain the optical, UV, and X-ray properties of the event.

Interestingly, we observed the electron energy index determined from the optically thin part of the synchrotron emission to fluctuate between $p\approx2.6$ to $p=3-3.5$ from 400\,d post-disruption on timescales of months. We deduce that these fluctuations could be due to either inhomogenous emitting regions in a spherical or conical outflow or fluctuations in the energy injection rate of a conical outflow. 
We rule out the possibility of a relativistic jet producing the outflow in this event since it would require an opening angle of the jet that is smaller than expected for a jet from an SMBH. We found that the mean speed of the outflow to $t = 183$~d was $\sim 0.1 c$, thus ruling out a long-lived relativistic outflow.  We also found a possible increase in deceleration, suggesting a transition to a Sedov-like evolution after $t \sim 600$~d.

We deduce that the unbound debris stream is unlikely to explain the radio properties of this event due to the high energy inferred through an equipartition analysis of the synchrotron emission. We propose the outflow could have originated from an accretion-driven wind or sub-relativistic jet, or a collision induced outflow from stream-stream intersections of the tidal debris.
Further detailed modelling of the multiwavelength properties of this event is required to differentiate between outflow models and explain some of the unique properties observed. Future observations of the radio decay of AT2019azh will enable more robust constraints on the CNM density and further insight into the nature of the outflow.   

\section*{Acknowledgements}
The authors thank K. Alexander for helpful discussions, W. Lu for providing helpful comments on an earlier version of this manuscript, and N. Blagorodnova, S. Kulkarni, A. de Witt, B. Cenko, S. Gezari, R. Fender, P. Woudt, and M. Bottcher for their contributions. This work was supported by the Australian government through the Australian Research Council's Discovery Projects funding scheme (DP200102471). The ZTF forced-photometry service was funded under the Heising-Simons Foundation grant $\#$12540303 (PI: Graham). The MeerKAT telescope is operated by the South African Radio Astronomy Observatory, which is a facility of the National Research Foundation, an agency of the Department of Science and Innovation. This work was carried out in part using facilities and data processing pipelines developed at the Inter-University Institute for Data Intensive Astronomy (IDIA). IDIA is a partnership of the Universities of Cape Town, of the Western Cape and of Pretoria.

\section*{Data Availability}
The spectral fitting and equipartition modelling software written for the purpose of this work is publicly available on Github at \url{https://github.com/adellej/tde_spectra_fit}. The observations presented in Table 1 will be published online in machine-readable format at \textbf{url to be entered on publication}. 

\section*{Software}
This research made use of Matplotlib, a community-developed \texttt{Python} package \citep{Hunter2007}, NASA's Astrophysics Data System Bibliographic Services, the Common Astronomy Software Application package \texttt{CASA} \citep{McMullin2007}, the $w$-stacking CLEAN imager \citep{offringa-wsclean-2014}, and the \texttt{Python} packages cmasher \citep{cmasher}, and emcee \citep{emcee}.


\bibliographystyle{mnras}
\bibliography{bibfile}

%


\appendix
\section{Consistency of radio observations across epochs}\label{sec:appendix1}
\begin{figure}
    \centering
    \includegraphics[width=0.49\textwidth]{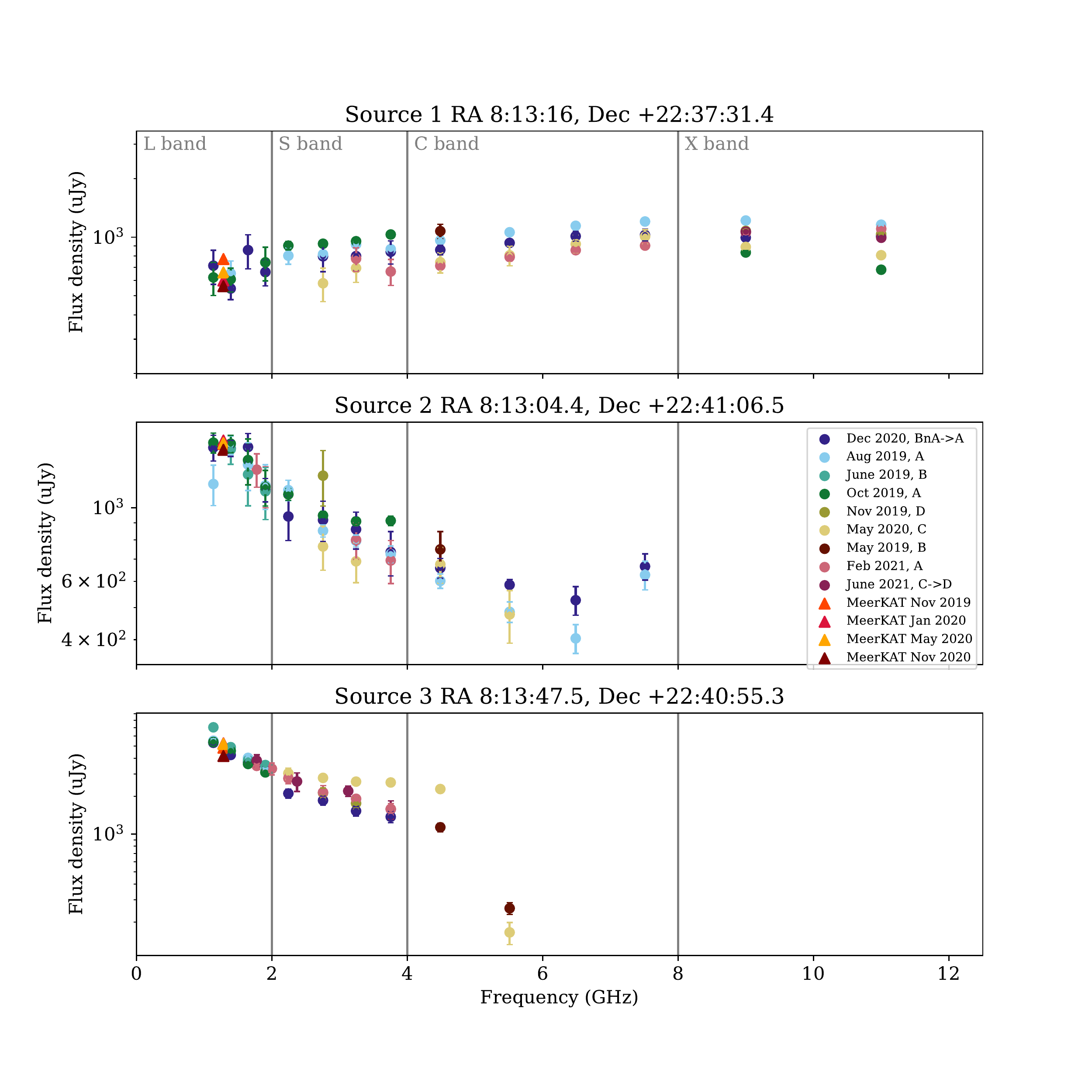}
    \caption{Flux density measurements of 3 background sources for 8 epochs of observations including VLA and MeerKAT data. The data indicate that no systematic offset is present between epochs or between instruments, but that fluctuations on the order of $\sim10\%$ are present between epochs of VLA observations.}
    \label{fig:bgsources}
\end{figure}

Due to the use of both the MeerKAT and VLA telescopes, and different configurations at the latter, we performed a consistency check of the data to ensure there were no instrumental or systematic offsets between epochs and between the VLA and MeerKAT L-band observations. We measured the flux densities of 3 background sources in the field of view for 9 of the VLA epochs and four of the MeerKAT epochs. We applied a primary beam correction to the images before extracting the flux densities. The flux densities for these 3 sources in all bands are shown in Figure \ref{fig:bgsources}, demonstrating that there is no significant systematic offset between the two instruments. 

In Figure \ref{fig:bgsources} there is evidence for fluctuations between VLA epochs on the order of $\sim10\%$. We examined the flux density scale obtained during calibrations of the secondary calibrator for the VLA observations. We found fluctuations of only a few percent between epochs, consistent with the expected flux-density calibration errors. Thus, the fluctuations between epochs are larger than expected due to flux calibration alone, and may be attributed to either intrinsic fluctuations in the sources we examined, primary beam corrections, resolution changes from different configurations, or other calibration inconsistencies. 

In the spectral observations of AT2019azh we identify steepenings of the synchrotron energy index. Here we rule out the possibility of these being artificial steepenings due to inconsistencies in the data calibration between epochs. In Figure \ref{fig:bgsources} there is no evidence for inconsistent calibration with frequency for single epochs, except the October 2019 epoch. If the steepenings were artificial, we would expect to see the epochs where the steepenings were observed to show a trend of being lower at the higher frequencies and higher at the lower frequencies than the epochs in which no steepenings were observed. The only epoch where evidence of this frequency trend is present is the October 2019 epoch, and thus we deduce that the slight steepening observed in this epoch (from $p\approx2.7$ to $p\approx3$) is likely not real and merely an artefact of the data calibration. For the other epochs, May 2020 and Feb 2021, there is no evidence in Figure \ref{fig:bgsources} of any such trend with frequency, and we deduce that the spectral steepenings we observed are real. 

The flux density offsets between epochs present in Figure \ref{fig:bgsources} would only affect the peak flux densities that we infer from the spectra, and not the spectral slope, since the offsets are not frequency dependent within epochs. Since the peak flux density is often not well constrained in our observations, the uncertainty in the peak flux density is dominated by the spectral fit uncertainty, and the fluctuations of order $\sim10\%$ between epochs are outweighed.

\section{Synchrotron emission fits including a cooling break}\label{sec:appendix2}
Recently \citet{Cendes2021} detected evidence for a cooling break in radio observations of the synchrotron emission from AT2019dsg. Here we analyse whether a cooling break is detected in our radio observations of AT2019azh. In the case of a cooling break, the data would indicate an additional steepening at a frequency $\nu_c > \nu_a$, and equation \ref{eq:spectralfit} is multiplied by 

\begin{equation}
    \left[ 1 + \frac{\nu}{\nu_c}^{s_2(\beta_2-\beta_3)} \right]^{-1/s_2}
\end{equation}
where $\beta_3$ = -$p$/2 and $s_2$ is a softening factor \citep{Granot2002}. 

To determine if there is a clear preference for a model with or without a cooling break, we compute the Akaike Information Criterion (AIC) and the Bayesian Information Criterion (BIC) of a spectral fit with a cooling break and one without. In both cases, the best model is indicated by the one with the smallest AIC or BIC. The AIC selects the best predictive model among a number of possibly misspecified models and is given by

\begin{equation}
    \mathrm{AIC}(M_k) = -2l_k \theta_k + 2p_k
\end{equation}
where $M_k$ is the model under consideration, $l_k$ is the log-likelihood of the model given its parameters $\theta_k$, and $p_k$ is the number of parameters estimated by the model $M_k$. 

The BIC selects the true model, using a minimal number of parameters and sets a large penalty for models with a larger set of parameters to prevent over-fitting. The BIC is given by

\begin{equation}
    \mathrm{BIC}(M_k) = -2l_k(\theta_k) + \ln{(n)}p_k
\end{equation}
where $n$ is the total number of data points that the model is being fit to. 

Here we define one model being significantly better than other if the preferred model has an AIC or BIC score of at least 2 units lower. A score of 0--2 means minimal confidence, a score of 2--6 means positive confidence, and a score $>$6 indicates strong confidence that the model is preferred. 

We carried out the same MCMC fitting as described in Section \ref{sec:Observations} but with the inclusion of the cooling break term in the model for epochs $t>321$\,d. When the additional cooling break was included in the fits, we found that the uncertainties on the peak frequency, peak flux, and $p$ increased, and the cooling break was identified to be around 4\,GHz. In Table \ref{tab:coolingbreak} we show the calculated AIC and BIC for the original model (model 1) and the model including a cooling break (model 2) for the epochs fit. 

\begin{table}
    \centering
    \caption{AIC and BIC values for the observed flux spectra for each epoch fit with the original model (model 1) and the model including a cooling break (model 2).}
    \begin{tabular}{c|cccccc}
    	\hline
		$\delta t$ (d) & 296 & 350 & 459 & 666 & 749 & 849 \\
		\hline
		\hline
		AIC$_1$ & 87.5 & 84.0 & 133.9 & 170.8 & 128.0 & 107.0\\
		AIC$_2$ & 89.9 & 85.9 & 135.7 & 169.2 & 132.9 & 110.9 \\
		\hline
		BIC$_1$ & 87.8 & 84.4 & 135.8 & 173.6 & 130.2 & 108.2 \\
		BIC$_2$ & 90.2 & 86.3 & 138.1 & 172.7 & 135.7 & 112.4 \\
        \hline
    \end{tabular}

    \label{tab:coolingbreak}
\end{table}

In Table \ref{tab:coolingbreak} it is clear that the AIC and BIC for model 1 is always lower than for model 2, with the model without a cooling break clearly preferred for the epoch at 850\,d, and marginally preferred for the other epochs. In this analysis we do not conclusively detect the presence of a cooling break. 

\bsp	
\label{lastpage}
\end{document}